\UseRawInputEncoding
\documentclass[aip,graphicx]{revtex4-1}
\usepackage{amsthm}
\usepackage{mathrsfs}
\usepackage{amssymb}
\usepackage{epsfig}
\usepackage{amsmath}
\usepackage{graphicx}
\usepackage{epsfig}
\usepackage{booktabs}
\usepackage{multirow}
\usepackage[colorlinks,urlcolor=blue]{hyperref}
\usepackage{makecell}
\usepackage{bm}
\usepackage[utf8]{inputenc}
\usepackage{lipsum}
\usepackage{soul}
\soulregister\cite7

\begin{document}

\author{Zhong Shen}
\author{Yufei Xue}
\author{Zebin Wu}
\address{Key Laboratory of Optical Field Manipulation of Zhejiang Province, Department of Physics, Zhejiang Sci-Tech University, Hangzhou 310018, China}

\author{Changsheng Song}
\email{cssong@zstu.edu.cn}

\address{Key Laboratory of Optical Field Manipulation of Zhejiang Province, Department of Physics, Zhejiang Sci-Tech University, Hangzhou 310018, China}
\address{Longgang Institute of Zhejiang Sci-Tech University, Wenzhou 325802, China}

\title{Enhanced Curie temperature and skyrmion stability in room temperature ferromagnetic semiconductor CrISe monolayer}


\begin{abstract}
We report CrISe monolayer as a room temperature ferromagnetic semiconductor with the Curie temperature ($T_C$), magnetic anisotropy energy (MAE) and band gap being 322 K, 113 $\mu$eV and 0.67 eV, respectively. The $T_C$ and MAE can be further enhanced up to 385 K and 313 $\mu$eV by tensile strain. More interestingly, the magnetic easy axis can be switched between off-plane and in-plane by compressive strain. Particularly, due to the broken inversion symmetry and strong spin-orbit coupling of Se atoms, a large Dzyaloshinskii-Moriya interaction (DMI) of 2.40 meV is obtained. More importantly, by micromagnetic simulations, stable skyrmions with sub-10 nm radius are stabilized by the large DMI above room temperature in a wide range of strain from $-2\%$ to $6\%$. Our work demonstrates CrISe as a promising candidate for next-generation skyrmion-based information storage devices and provides guidance for the research of DMI and skyrmions in room temperature ferromagnetic semiconductors.
\end{abstract}
\maketitle
\section{Introduction}
Since the long-range magnetic order is discovered in monolayer CrI$_3$\cite{Mcguire2015,Huang2017b,RN15} and bilayer Cr$_2$Ge$_2$Te$_6$\cite{Hao2018,Wang2018,Gong2017}, two-dimensional (2D) van der Waals (vdW) magnetic materials have attracted more and more attention. However, the low Curie temperature ($T_C$) below 70K greatly limits their practical applications. A plenty of efforts\cite{Tu2016,Furdyna1988,Dalpian2006,Pan2008,Sato2010a,Dietl2010} have been made to enhance the $T_C$ of semiconductors. However, confined by the dopant-host hybridization\cite{Dalpian2006,Wei2008}, the highest $T_C$ is reported only $\sim $200 K\cite{Hao2018,Bouzerar2007,Chen2011,Zutic2018}. Although room-temperature ferromagnetic (FM) metallic monolayers such as VSe$_2$\cite{Bonilla2018,Liu2018b} and Fe$_3$GeTe$_2$\cite{Tan2018,Deng2018,Fei2018} are experimentally exfoliated, 2D FM semiconductors with high $T_C$ above room temperature are rather scarce and yet to be further researched\cite{Wang2019d}.

Recent years, skyrmion\cite{RN2009,RN120,RN121,RN109} has been one of the most eye-catching fields of 2D FM materials due to its promising application in next-generation information storage\cite{RN887,RN1986,RN1988,RN116}. With distinctive topological protection, skyrmion can keep high stability even at small size\cite{RN3121,RN3542,RN1040,RN3177} and it's less susceptible to defects in the crystal. Meanwhile, Dzyaloshinskii-Moriya interaction (DMI)\cite{RN130,RN131,RN121,Nagaosa2013} that plays a vital role in the creation and stability of skyrmions has been widely investigated\cite{RN24,RN33,RN32,RN2001}.  Since the inversion symmetry of most 2D magnetic materials, DMI are absent in them. While the Janus structure\cite{RN30,RN31,RN1041,RN1059,PhysRevB.101.094420,RN3528} with intrinsic inversion asymmetry is reported as a promising way to induce large DMI. However, such Janus monolayers are rather scarce especially the one that with high $T_C$ above room temperature.

Here, by first-principles calculations and micromagnetic simulations, we demonstrate CrISe monolayer as a room-temperature FM semiconductor with an indirect band gap of 0.67 eV and an off-plane magnetic anisotropy energy (MAE) of 113 $\mu $eV. Then, the intrinsic large DMI of 2.40 meV is investigated which can stabilize skyrmions without any external magnetic field. In addition, a high $T_C$ of 322 K is found in its intrinsic configuration, and then, the $T_C$ and MAE can be further enhanced up to 385 K and 313 $\mu $eV under tensile strain, respectively. Besides, DMI monotonically increases with the maximum value reaching 3.28 meV. The micro-mechanism of DMI is also explored from the view of atom- and orbit-resolved spin-orbit coupling (SOC) energy difference. The results show that the strong SOC of Se atoms plays a leading role. More importantly, room temperature skyrmions can maintain stability within a wide strain range from $-2\%$ to $6\%$ and a skyrmion phase can be induced by a suitable external magnetic field. The integration of high $T_C$, intrinsic semiconductor properties, large DMI and skyrmions in CrISe monolayer makes it a promising candidate for spintronic applications.

\section{Methods}
\label{II}

The first-principles calculations are performed by the Vienna ab initio simulation package (VASP)\cite{RN2004} with density-functional theory (DFT). 
The projected augmented wave (PAW) method\cite{RN1060,RN1064,RN1061} is chosen to describe the electron-core interaction with the generalized gradient approximation (GGA) of Perdew-Burke-Ernzerhof (PBE)\cite{RN1999} being employed for the exchange correlation functional. The DFT+U
method\cite{Dudarev1998a} is carried out to treat the electron correlations between 3$d$ electrons of Cr with an effective Hubbard-like term U = 1.5 eV. The primitive cell is fully relaxed with a high convergence standard of energy and force less than 10$^{-7}$ eV and $-$0.01 eV/\AA{}, respectively. The plane wave cutoff energy is set as 500 eV and a 6 $\times$ 6 $\times$ 1 $\Gamma$-centered k-point mesh is used to carry out the first Brillouin-zone integration. Besides, to exclude the interaction between adjacent layers, a 15 \AA{} vacuum space is adopted along the Z axis. The phonon spectrum is calculated by using the PHONOPY code\cite{Togo2008,Togo2015} with a 3 $\times$ 3 $\times$ 1 supercell.

The micromagnetic simulations are performed by the \textit{Spirit} package\cite{RN1055} with Landau-Lifshitz-Gilbert (LLG) equation\cite{RN1984,RN1985}.
In the simulations, we use a supercell that contains 32400 sites with periodic boundary conditions and random initial state of spins. The $T_C$ of CrISe monolayer is evaluated by using the MCSOLVER code\cite{Liu2019} with
Monte Carlo (MC) simulations and the Metropolis algorithm. To sufficiently relax the system, we employ 4 $\times$ 10$^4$ sweeps for thermalization at each temperature point and 2 $\times$ 10$^5$ steps are used for statistics of observables. 

To explore the magnetic properties of Janus CrISe, we use the following spin Hamiltonian:
\begin{equation}\label{1}
	H = \sum\limits_{\left \langle i,j \right \rangle}J_{1}(\vec{S_{i}}\cdot\vec{S_{j}}) + \sum\limits_{\left \langle k,l \right \rangle}J_{2}(\vec{S_{k}}\cdot\vec{S_{l}}) + \sum\limits_{\left \langle m,n \right \rangle}J_{3}(\vec{S_{m}}\cdot\vec{S_{n}}) + \sum\limits_{\left \langle i,j \right \rangle}\vec{d_{ij}}\cdot(\vec{S_{i}}\times\vec{S_{j}}) + 
K\sum_{i}(S_{i}^{z})^2 + \mu_{Cr}B\sum_{i}S_{i}^{z},
\end{equation}
where $J_{1}$, $J_{2}$, $J_{3}$ are the Heisenberg exchange coefficients between nearest-neighbor (NN), second NN, third NN Cr atoms and $\vec{d_{ij}}$ is the DMI vectors between spin $\vec{S_i}$ and $\vec{S_j}$. $K$ represents the single ion anisotropy and $S_i^z$ is the $z$ component of $\vec{S_i}$. $\mu_{Cr}$ and $B$ are the magnetic moment of Cr atom and external magnetic field, respectively.

To obtain the Heisenberg exchange coefficients ($J_1$, $J_2$, $J_3$) and DMI vector ($\vec{d}$), we employ the widely used total energy difference approach\cite{RN33,RN32,RN30}. $J_1$, $J_2$, $J_3$ and $d_{//}$ can be expressed as:
\begin{equation}\label{2}
	J_1 = \frac{E_{1}}{36} - \frac{E_{2}}{36} - \frac{E_{3}}{12} + \frac{E_{4}}{12} .
\end{equation}
\begin{equation}\label{3}
	J_2 = \frac{E_{1}}{72} - \frac{E_{2}}{72} + \frac{E_{3}}{12} - \frac{E_{4}}{12} .
\end{equation}
\begin{equation}\label{4}
	J_3 = \frac{E_{1}}{144} + \frac{E_{2}}{18} - \frac{E_{3}}{48} - \frac{E_{4}}{24} .
\end{equation}
\begin{equation}\label{5}
	d_{//} = (E_{CW} - E_{ACW})/12 .
\end{equation}
Here, $E_1$, $E_2$, $E_3$, $E_4$, $E_{CW}$ and $E_{ACW}$ are the total energies of the system with different spin configurations (see supplemental material for more details).  
Then, according to Moriya's symmetry rules\cite{RN131}, the DMI vector between each pair of the nearest-neighbor Cr atoms can be expressed as $\vec{d_{ij}} = d_{//}(\vec{u_{ij}}\times\vec{z}) + d_{z}\vec{z}$\cite{RN2,shen2021}. Here, $\vec{u_{ij}}$ is the unit vector between sites $i$ and $j$ and $\vec{z}$ is the unit vector along $Z$ direction. $d_{//}$ and $d_z$ are the in-plane and off-plane components of $\vec{d_{ij}}$ with $d_{z}$ = $\frac{d_{//}}{\tan\alpha}$ \cite{RN30,RN2}. Here, $\alpha$ is the average of the tilting angle of the atomic plane (Cr)$_i$-I(Se)-(Cr)$_j$ (see Fig. S2(a) of supplemental material).

\section{Results and discussion}
\label{III}

\begin{figure}[!htbp] 
	\begin{center}
		\includegraphics[width=14.0cm]{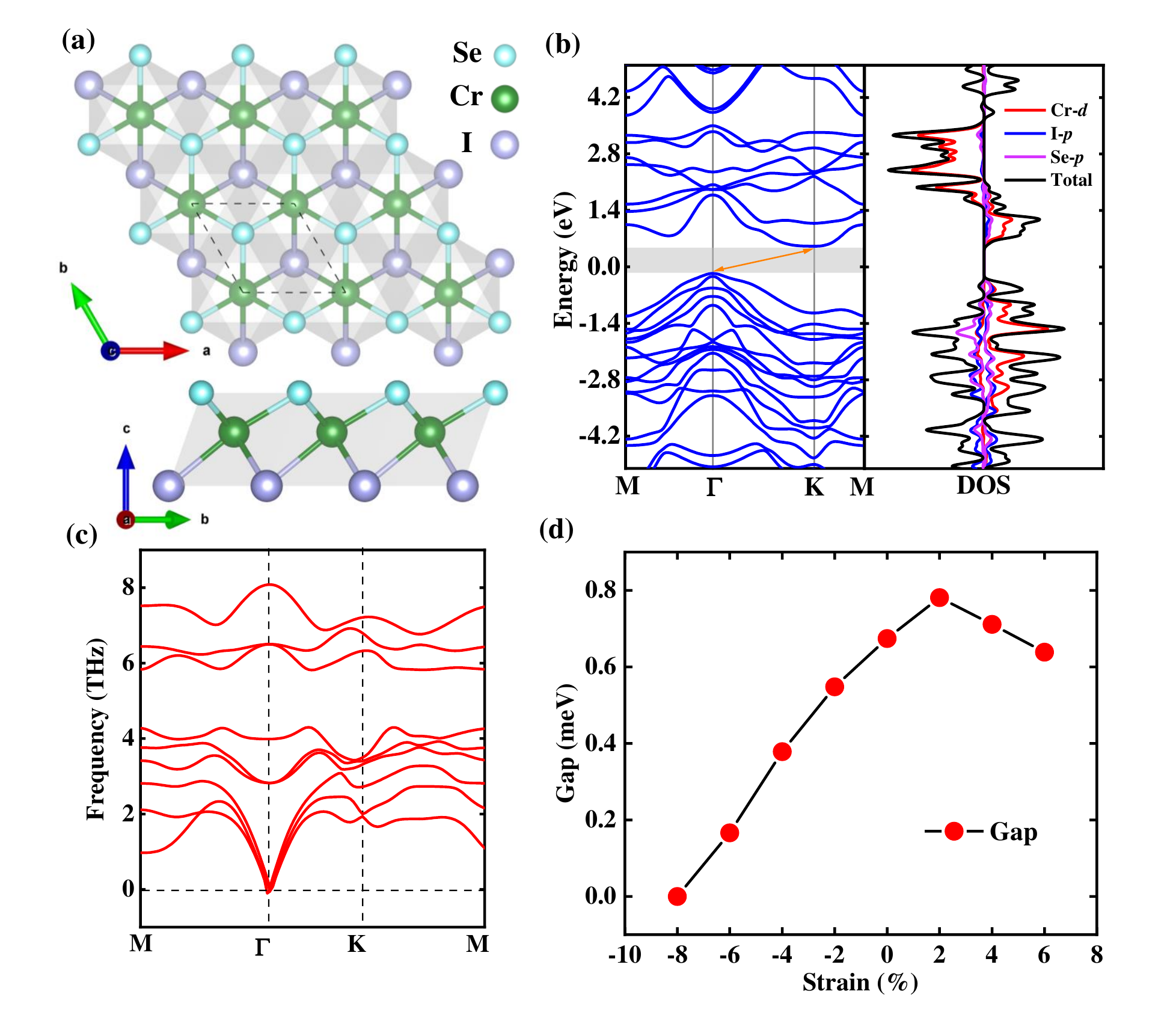}\
		\caption{\label{fig1}
			(a) Top and side views of ball-stick structure, (b) Band structures and density of states (DOS), (c)  Phonon spectrum along the high symmetry path in reciprocal space, and (d)	Band gap as a function of biaxial strain in CrISe monolayer.}
\end{center}
\end{figure}
The top and side views of Janus CrISe monolayer are shown in Fig. \ref{fig1}(a) with the cyan, purple and green balls representing I, Se and Cr, respectively. Cr atoms with point group $C_{3v}$ form a hexagonal close-packed lattice (HCP) and are sandwiched by two nonmagnetic atomic planes consisting of I and Se atoms.
This special Janus structure makes CrISe monolayer possess intrinsic spatial inversion asymmetry along Z direction. The dashed lines in Fig. \ref{fig1}(a) show the unit cell of CrISe with the optimized lattice parameters $a = b = 3.747$ \AA{}. To see the stability of CrISe monolayer, we calculated the phonon spectrum along high symmetry points in reciprocal space as shown in Fig. \ref{fig1}(c). No imaginary frequency is found which suggests the dynamical stability of CrISe monolayer.

Figure. \ref{fig1}(b) shows the band structure and density of states (DOS) of CrISe monolayer. An indirect band gap of 0.67 eV is indicated by the orange arrow and gray shade in the band structure as well as the discontinuous region of DOS near the Fermi level. The band gap of CrISe as a function of biaxial strain is shown in Fig. \ref{fig1}(d). Here, the strain is defined as $\frac{a-a_{0}}{a_{0}}$ where $a$ and $a_0$ are the strained and unstrained lattice constant. It can be seen that CrISe monolayer is a semiconductor with non-zero band gap in the range of $-$6$\%$ to 6$\%$ biaxial strain. Then, the band gap drops to $\sim$ 0 meV and undergoes a metal behavior after applying a compressive strain larger than $-8\%$. (Band structures with different strains can be found in Fig. S3 of supplemental material.)

\begin{figure}[!htbp]
	\begin{center}
		\centering
		\includegraphics[width=14.0cm]{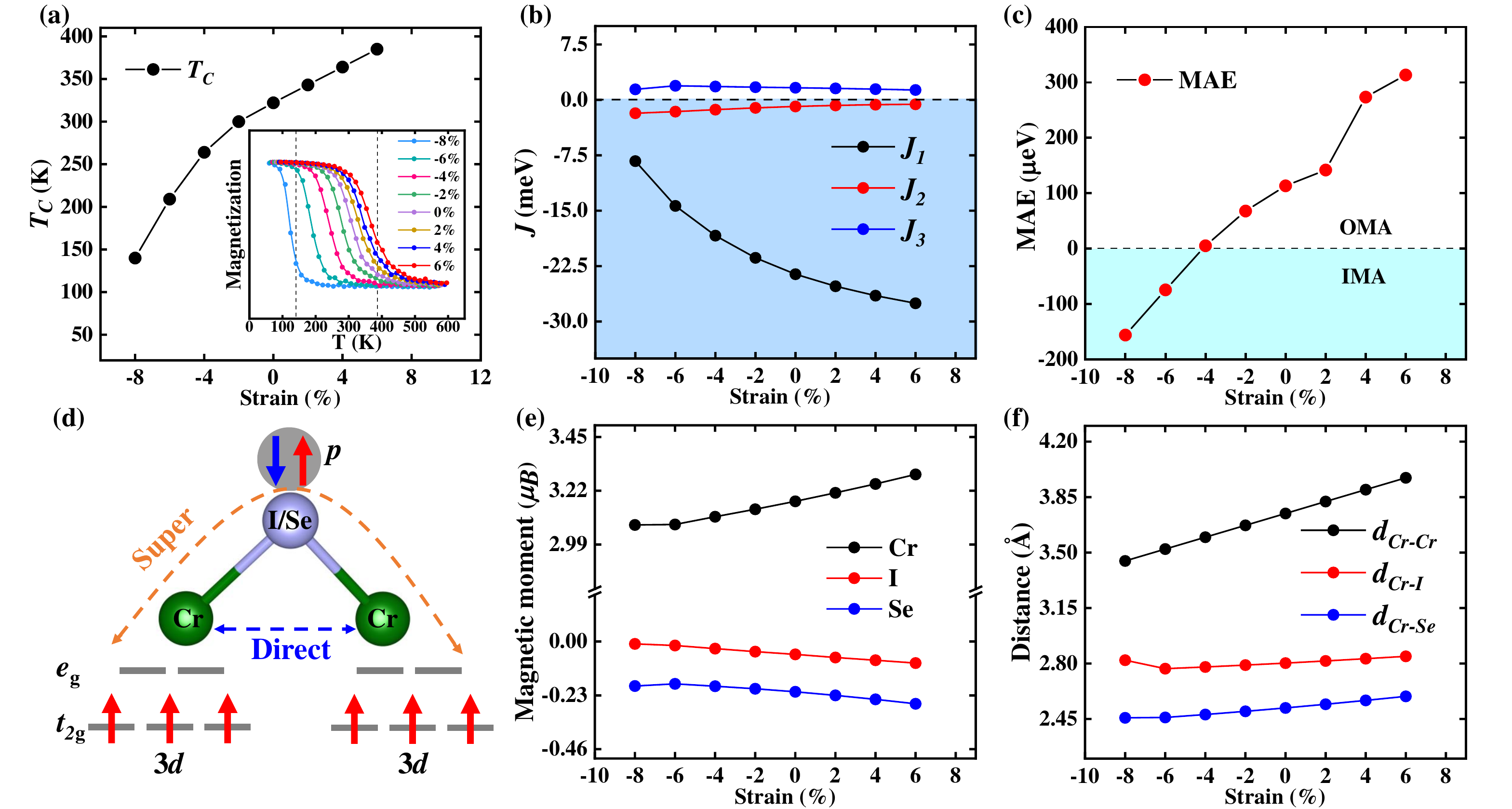}\
		\caption{\label{fig2}
			(a) Curie temperature ($T_{C}$) of CrISe as a function of biaxial strain. The inset of (a) demonstrates the magnetization as a function of temperature under different strains. (b) Heisenberg exchange coefficients ($J_1$, $J_2$, $J_3$) and (c) Magnetic anisotropy energy (MAE) of CrISe as functions of biaxial strain. (d) Schematic diagram of super and direct exchange interaction with the red (blue) arrows representing the spin up (down) electrons of $d$ or $p$ orbits. (e) Magnetic moments of Cr, I, Se ions and (f) bond lengths of Cr-Cr, Cr-I and Cr-Se as functions of biaxial strain.
				}
	\end{center}
\end{figure}

After discussing the geometry structure and dynamical stability of CrISe monolayer, we now focus on the Curie temperature ($T_C$) and magnetic anisotropy energy (MAE), which are crucial for long-range ferromagnetic order in 2D ferromagnets. As shown in Fig. \ref{fig2}(a), the $T_C$ of intrinsic CrISe is 322 K (above room temperature) without any biaxial strain. Then, by applying tensile strain, the $T_C$ can be further enhanced up to 385 K. Notice that the highest $T_C$ of ferromagnetic semiconductors is reported only $\sim$ 200 K\cite{Sato2010a,Bouzerar2007,Chen2011,Zutic2018,Wang2019d}. Such high $T_C$ of CrISe monolayer makes it a promising platform for nanoscale spintronic applications\cite{Guo,Li2019}. The insert of Fig. \ref{fig2}(a) shows the normalized magnetization as functions of temperature with different strain to evaluate the $T_C$. Then, the first, second and third nearest neighbor Heisenberg exchange coefficients $J_1$, $J_2$ and $J_3$ are shown in Fig. \ref{fig2}(b). $J_1$ and $J_2$ are negative that correspond to ferromagnetic (FM) coupling while the positive value of $J_3$ represents the antiferromagnetic (AFM) coupling. It's clearly seen that the values of $J_1$ is much larger than those of $J_2$ and $J_3$ suggesting that $J_1$ plays a leading role in determining the $T_C$ of CrISe.

The monotonically increasing magnitude of $J_1$ induced by biaxial strain is responsible to the enhanced ferromagnetism and the improved $T_C$.
To further explore the microphysical mechanisms of increasing $J_1$, as shown in Fig. \ref{fig2}(d), we investigate the variation of super ($J_S$) and direct ($J_D$) exchange interactions with $J_1 = J_D + J_S$ and the value of $J_1$ is determined by the competition between $J_S$ and $J_D$. The direct exchange interaction $J_D$ is positive that corresponds to a weak AFM coupling between the localized $3d$ electrons of nearest Cr pairs. According to Goodenough-Kanamori-Anderson (GKA) rules \cite{Goodenough1955,Kanamori1959,Anderson1959}, since the Cr-I(Se)-Cr bond angle is close to 90$^{\circ}$, the super exchange interaction ($J_S$) mediated by the $p$ electrons of I(Se) atom induces a FM coupling ($J_S < 0$).
Figure. \ref{fig2}(e) shows that the magnetic moments of Cr and I (Se) have opposite signs but their magnitudes all increase with strain in the range of $-6\%$ to $6\%$, which is consistent with the enhanced $T_C$. Besides, as illustrated in Fig. \ref{fig2}(f), the sharply increased bond length of Cr-Cr ($d_{Cr-Cr}$) corresponding to the reduced AFM direct exchange ($J_D$), while the bond lengths of Cr-I ($d_{Cr-I}$) and Cr-Se ($d_{Cr-Se}$) hardly change and maintain a strong FM super exchange ($J_S$). Thus, the magnitude of $J_1$ ($J_1 = J_D + J_S$) keeps increasing with strain in the range of $-8\%$ to $6\%$ which leads to the enhanced $T_C$.

Then, we focus on the magnetic anisotropy energy\cite{Lado2017,Lado2020SA} (MAE) of CrISe, which is defined as $E_{xx} - E_{zz}$. Here, $E_{xx}$ ($E_{zz}$) is the energy of the unit cell when spin points along X (Z) direction. As demonstrated in Fig. \ref{fig2}(c), the MAE increases with tensile strain but decreases with compressive strain. Then, a switch from off-plane magnetic anisotropy (OMA) to in-plane magnetic anisotropy (IMA) is observed when the compressive strain increase to $-6\%$. The microscopic physical mechanism of this interesting switch is explored and we find that the hybrid $p_x$-$p_z$ orbits of I atom is responsible to the IMA at $-6\%$ strain (more detailed explanations of the MAE shift can be found in Fig. S4 of supplemental material).

\begin{figure}[!htbp]
	\begin{center}
		\includegraphics[width=14.0cm]{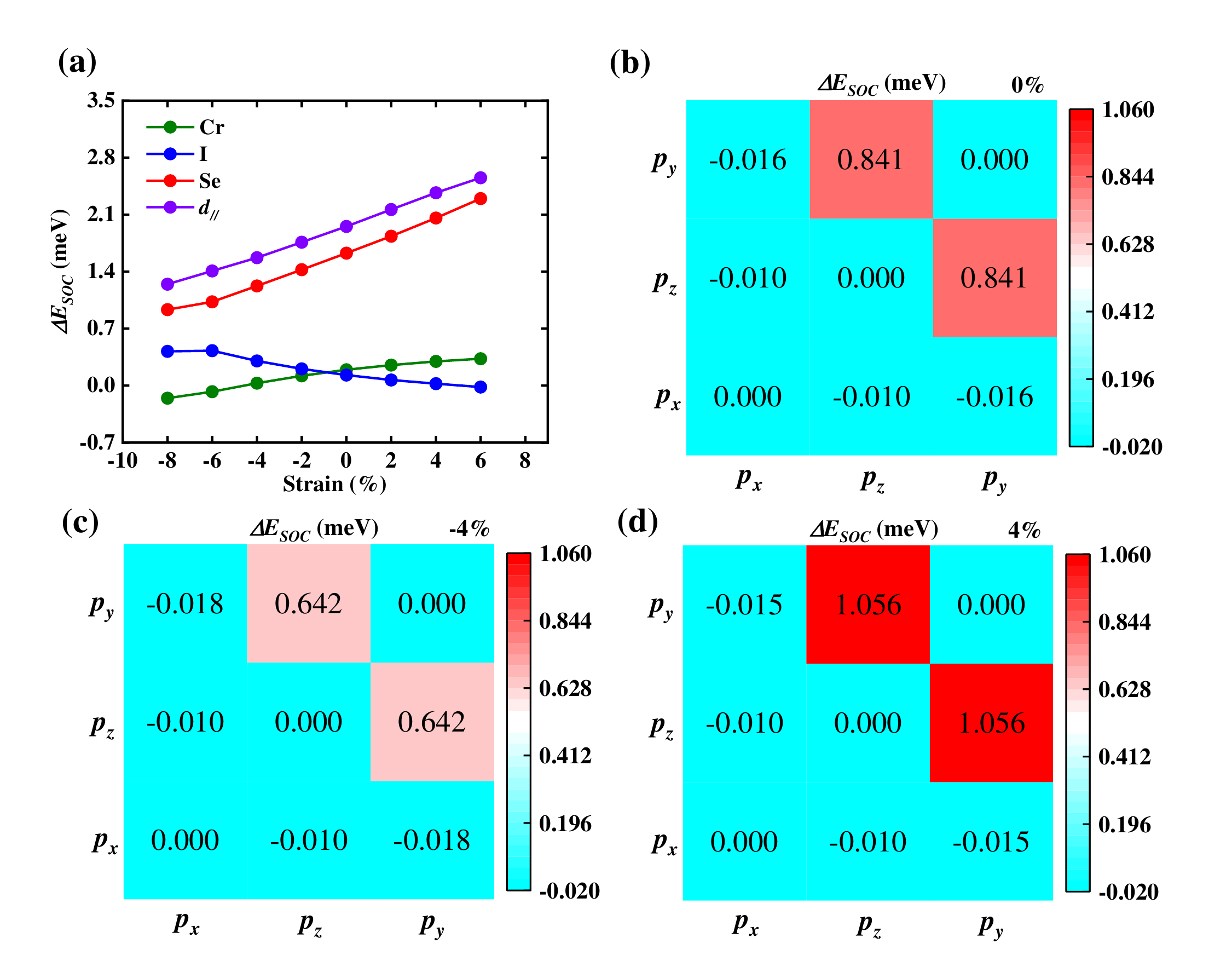}\
		\caption{\label{fig3}
			(a) Atom-resolved $\Delta E_{SOC}$ and in-plane DMI component ($d_{//}$) as functions of biaxial strain in CrISe monolayer. Orbit-resolved $\Delta E_{SOC}$ of Se-$p$ orbits with biaxial strain of (b) 0$\%$, (c) $-4\%$ and (d) 4$\%$.
		}
	\end{center}
\end{figure}

As mentioned above, we have demonstrated CrISe as a room temperature FM semiconductor with high $T_C$ of 322 K. The $T_C$ and band gap of CrISe are highly sensitive to biaxial strain and the MAE undergoes an easy-axis switch. Next, considering the intrinsic spatial inversion asymmetry, we calculate the Dzyaloshinskii-Moriya interaction (DMI) of CrISe as a function of biaxial strain as shown in Fig. \ref{fig3}(a) ($d$ = $\sqrt{d_{//}^2 + d_z^2}$ and $d_z$ are shown in the supplemental material). A strong DMI of 2.40 meV is found in intrinsic CrISe monolayer with the in-plane (off-plane) component $d_{//}$ ($d_z$) being 1.95 (1.40) meV. Besides, the DMI of CrISe increases from 1.45 meV to 3.28 meV monotonically in the range of $-8\%$ to $6\%$ biaxial strain. Then, to reveal the microscopic origin of DMI, we calculate the DMI associated SOC energy $\Delta E_{SOC}$ (the energy difference between CW and ACW spin configurations as shown in Fig. S2(a) of supplemental material) of different atoms as shown in Fig. \ref{fig3}(a). The dominant contribution to DMI comes from the heavy chalcogen Se atoms. This is similar to that of Cr$_2$X$_3$Y$_3$ monolayer\cite{shen2021} and ferromagnet/heavy metal heterostructures\cite{RN33} and it can be understood by the Fert-Levy mechanism that the spin-orbit scattering required by DMI is induced by the heavy chalcogen atoms\cite{RN131,PhysRevLett.44.1538}. To further elucidate the mechanisms of the increase (decrease) of the $\Delta E_{SOC}$ of Se atoms at tensile (compressive) strain, we calculate the contributions from 4$p$ orbitals hybridization of Se atoms with biaxial strains of $0\%$, $-4\%$ and $4\%$, respectively. As shown in Fig. \ref{fig3}(b), when no strain is applied, the $\Delta E_{SOC}$ of the hybridization between $p_y$ and $p_z$ is 0.841 meV and plays a leading role in determining the $\Delta E_{SOC}$ of Se atom compared to that of $p_x$-$p_y$ and $p_x$-$p_z$ hybridization. Then, when a biaxial strain of $4\%$ ($-4\%$) is applied, the $\Delta E_{SOC}$ of $p_y$ and $p_z$ hybridization increase (decrease) to 1.06 (0.642) meV, which explains the increased (reduced) DMI of Se atom with tensile (compressive) strain.

\begin{figure}[!htbp]
	\begin{center}
		\includegraphics[width=14.0cm]{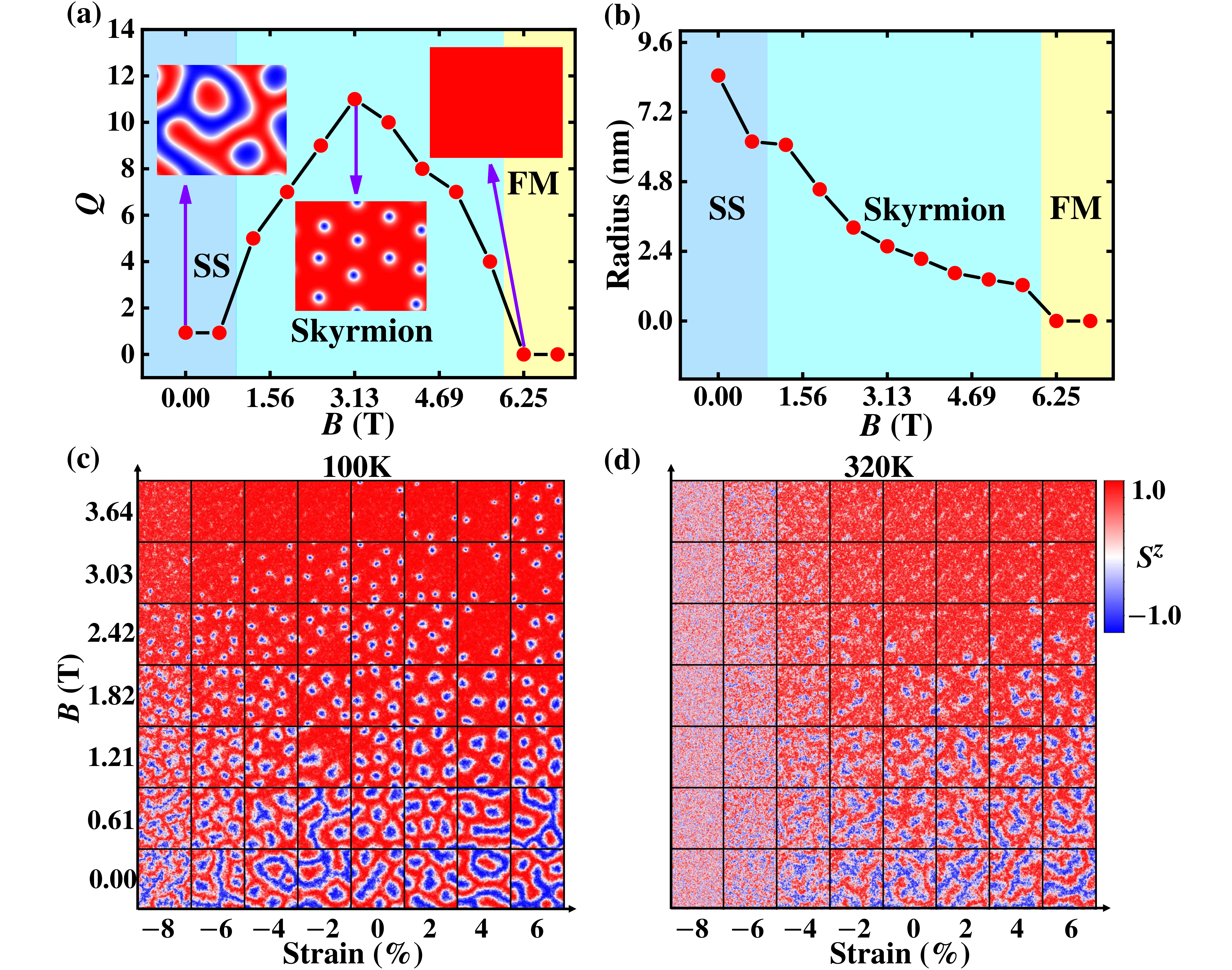}\
		\caption{\label{fig4}
		(a) Topological charge and (b) radius of skyrmion in CrISe supercell as functions of magnetic field ($B$) without strain at 0 K.
		Spin textures of CrISe with different magnetic fields and biaxial strains at (c) 100 and (d) 320 K. The color map indicates the off-plane spin component of Cr atoms.
		}
	\end{center}
\end{figure}

Next, we perform micromagnetic simulations to explore the spin textures of CrISe with the magnetic interaction parameters (Tab. S2) of Eq. \ref{1} obtained from first-principles calculations. The topological charge [$Q = \frac{1}{4\pi}\int \vec{S}\cdot (\partial_x \vec{S} \times \partial_y \vec{S})dxdy$\cite{Berg1981,PhysRevB.93.174403}] which reflects the density of skyrmions in the supercell is shown in Fig. \ref{fig4}(a). Without applying any magnetic field, skyrmions exist spontaneously accompanied by labyrinth domains forming the spin spiral phase (SS)\cite{RN3124,RN3705}. With the increase of magnetic field, the labyrinth domains gradually shrinks and splits into skyrmions, leading to the increased $Q$ at 0.63-3.13 T. Besides, the labyrinth domains completely disappear above 1.25 T and the skyrmion phase is formed with the typical image of isolated skyrmions in a ferromagnetic background. Such skyrmion phase is preserved within the range of 1.25-5.63 T but the density of skyrmions decreases in the range of 3.13-5.63 T. Then, if the applied magnetic field is larger than 5.63 T, the skyrmions will be annihilated into a FM state. Moreover, as shown in Fig. \ref{fig4}(b), the radius of skyrmion in intrinsic CrISe is sub-10nm, which is significantly smaller than those of MnSTe (20 nm)\cite{RN1042} and CrInSe$_3$ (15.1 nm)\cite{RN3705}. Then, by applying suitable magnetic field, the radius of skyrmion can be further regulated to 1.24 nm. Such small-size skyrmions, as potential information carriers, are urgently desired for future high-density data storage and information processing devices\cite{RN3705,RN3663,RN3063}.

Then, considering that skyrmions in 2D magnets often exist only at low temperature\cite{RN3528,RN1059,RN1041,RN30} which prevents the application of room-temperature spintronic devices, we further study the effect of temperature on the spin textures of CrISe with different strains.
The spin textures at a temperature of 100 and 320 K are presented in Fig. \ref{fig4}(c) and \ref{fig4}(d).  At a temperature of 100 K, without the existence of magnetic field, the labyrinth domains can be clearly seen in the whole range of strain from $-8\%$ to $6\%$. Then, the skyrmion phase can be induced by applying an increasing magnetic field. However, as shown in Fig. \ref{fig4}(d), when the temperature increases to 320 K (above room temperature), the spin textures of CrISe with compressive strain change to paramagnetic state due to the sub-320 K $T_C$. On the contrary, thanks to the high $T_C$ (above 322 K) of CrISe under tensile strain from $0\%$ to $6\%$, isolated skyrmions are stabilized in such high temperature with suitable magnetic fields (0.61-2.42 T). In addition, similar to the case at 0 K (see Fig. \ref{fig4}(b)), the density of skyrmions increase firstly and then decreases with increasing magnetic fields. At last, a polarized FM state accompanied by local thermal fluctuations can be induced when the magnetic fields further increase as shown in Fig. \ref{fig4}(c) and \ref{fig4}(d).

\begin{figure}[!htbp]
	\begin{center}
		\includegraphics[width=14.0cm]{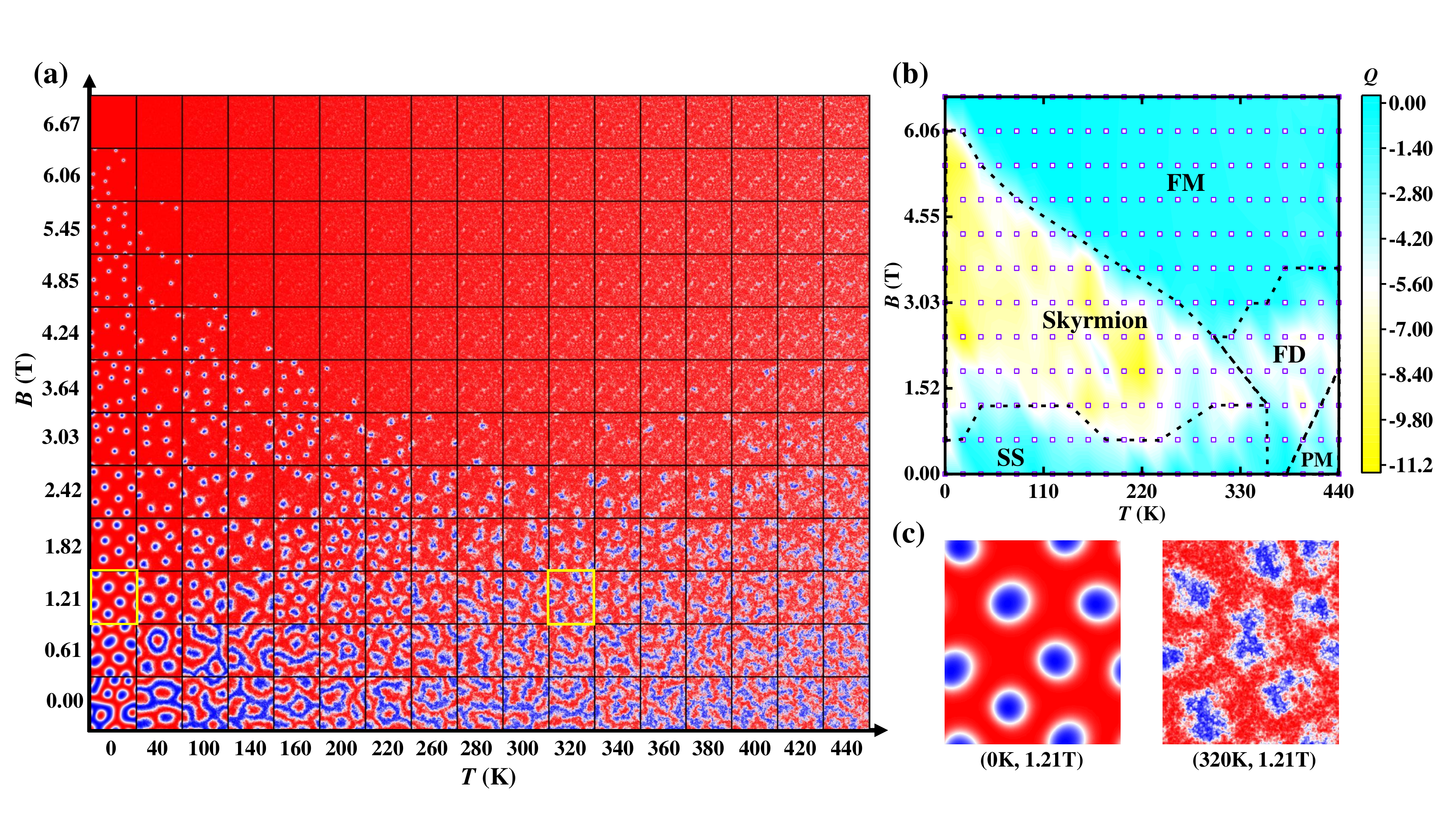}\
		\caption{\label{fig5}
		Evolutions of (a) spin textures and (b) topological charge of CrISe supercell with variational magnetic fields and temperatures. (c) The enlarged views of typical skyrmion phase at 0 and 320 K as outlined in (a) with yellow lines.}
	\end{center}
\end{figure}
Moreover, we further investigate the evolutions of spin textures and topological charge with changing magnetic fields and temperatures as shown in Fig. \ref{fig5}. In the micromagnetic simulations, we employ the magnetic parameters of CrISe with a tensile strain of 6$\%$ considering the high $T_C$ of 385 K. It can be seen from Fig. \ref{fig5}(a) and \ref{fig5}(b) that three phases named spin spiral (SS), skyrmion and FM appear sequentially with the increase of magnetic field when the temperature is lower than 300 K. Besides, as can be seen from Fig. \ref{fig5}(b), with temperature increasing, the skyrmion phase exists in a narrower range of magnetic field. This is similar to that of Cu$_2$OSeO$_3$\cite{RN3177} and MnBi$_2$(Se, Te)$_4$\cite{RN3528} systems. When the temperature increases above room temperature, a new phase named fluctuation-disorder (FD)\cite{RN3528,RN3705} appears. At this phase, skyrmions begin to be destroyed by the thermal fluctuations, but not completely annihilated. Moreover, the breaking of skyrmions creates more little vortices which slightly increases the topological charge $Q$ as can be seen from Fig. \ref{fig5}(b). Then, the FD phase can be modulated to the polarized FM phase by a strong magnetic field or turning to the PM phase with higher temperature.

\section{Conclusion}
\label{IV}
In conclusion, by first-principles calculations and micromagnetic simulations we demonstrate intrinsic CrISe monolayer as a room temperature ferromagnetic semiconductor with high Curie temperature ($T_C$) of 322 K and large band gap of 0.67 eV. Then, the $T_C$ can be further enhanced to 385 K by a 6$\%$
 tensile strain and the magnetic easy axis can be switched between in-plane and off-plane by compressive strain. Besides, a large Dzyaloshinskii-Moriya interaction (DMI) of 2.40 meV is found in intrinsic CrISe monolayer. The DMI monotonically increases in $-8\%$ to $6\%$ biaxial strain with the maximum value reaching 5.42 meV. Microscopically, the DMI is mainly associated with the strong spin-orbit coupling induced by Se atoms. Such large DMI can stabilize skyrmions with sub-10 nm radius when the magnetic field is absent. More importantly, the skyrmion phase is induced by suitable magnetic fields above room temperature in a wide range of strain. In addition, the temperature-magnetic field phase diagram is mapped out to investigate the effect of thermal fluctuations on the stability of skyrmions. Our findings demonstrate CrISe monolayer as a promising candidate for spintronic devices and provide beneficial guidance for designing next-generation skyrmion-based information storage devices.

\section{Supplemental Materials Available}
1. Calculation details of the Heisenberg exchange coefficients ($J_1$, $J_2$, $J_3$) and Dzyaloshinskii-Moriya interaction.
2. Atom-resolved MAE of CrISe monolayer and orbit-resolved MAE of I-$p$ orbits with biaxial strains of $-6\%$, $0\%$ and $6\%$.
3. Band structures of CrISe with different biaxial strains.
4. Detailed structure parameters of CrISe monolayer.
5. Specific values of Heisenberg exchange coefficients ($J_1$, $J_2$, $J_3$), in-plane ($d_{//}$) and off-plane ($d_z$)
components of DMI, single ion anisotropy ($K$) and the magnetic moments of Cr atoms with different strains.

\begin{acknowledgments}
	This work was supported by National Natural Science Foundation of China (Grant No.11804301), the Natural Science Foundation of Zhejiang Province (Grant No.LY21A040008), the Fundamental Research Funds of Zhejiang Sci-Tech University (Grant Nos.2021Q043-Y,LGYJY2021015).
\end{acknowledgments}

\bibliography{ref}

\begin{thebibliography}{76}%
\makeatletter
\providecommand \@ifxundefined [1]{%
 \@ifx{#1\undefined}
}%
\providecommand \@ifnum [1]{%
 \ifnum #1\expandafter \@firstoftwo
 \else \expandafter \@secondoftwo
 \fi
}%
\providecommand \@ifx [1]{%
 \ifx #1\expandafter \@firstoftwo
 \else \expandafter \@secondoftwo
 \fi
}%
\providecommand \natexlab [1]{#1}%
\providecommand \enquote  [1]{``#1''}%
\providecommand \bibnamefont  [1]{#1}%
\providecommand \bibfnamefont [1]{#1}%
\providecommand \citenamefont [1]{#1}%
\providecommand \href@noop [0]{\@secondoftwo}%
\providecommand \href [0]{\begingroup \@sanitize@url \@href}%
\providecommand \@href[1]{\@@startlink{#1}\@@href}%
\providecommand \@@href[1]{\endgroup#1\@@endlink}%
\providecommand \@sanitize@url [0]{\catcode `\\12\catcode `\$12\catcode
  `\&12\catcode `\#12\catcode `\^12\catcode `\_12\catcode `\%12\relax}%
\providecommand \@@startlink[1]{}%
\providecommand \@@endlink[0]{}%
\providecommand \url  [0]{\begingroup\@sanitize@url \@url }%
\providecommand \@url [1]{\endgroup\@href {#1}{\urlprefix }}%
\providecommand \urlprefix  [0]{URL }%
\providecommand \Eprint [0]{\href }%
\providecommand \doibase [0]{http://dx.doi.org/}%
\providecommand \selectlanguage [0]{\@gobble}%
\providecommand \bibinfo  [0]{\@secondoftwo}%
\providecommand \bibfield  [0]{\@secondoftwo}%
\providecommand \translation [1]{[#1]}%
\providecommand \BibitemOpen [0]{}%
\providecommand \bibitemStop [0]{}%
\providecommand \bibitemNoStop [0]{.\EOS\space}%
\providecommand \EOS [0]{\spacefactor3000\relax}%
\providecommand \BibitemShut  [1]{\csname bibitem#1\endcsname}%
\let\auto@bib@innerbib\@empty
\bibitem [{\citenamefont {Mcguire}\ \emph {et~al.}(2015)\citenamefont
  {Mcguire}, \citenamefont {Dixit},\ and\ \citenamefont
  {Cooper}}]{Mcguire2015}%
  \BibitemOpen
  \bibfield  {author} {\bibinfo {author} {\bibfnamefont {M.}~\bibnamefont
  {Mcguire}}, \bibinfo {author} {\bibfnamefont {H.}~\bibnamefont {Dixit}}, \
  and\ \bibinfo {author} {\bibfnamefont {V.}~\bibnamefont {Cooper}},\ }\href
  {\doibase 10.1021/cm504242t} {\bibfield  {journal} {\bibinfo  {journal}
  {Chem. Mater.}\ }\textbf {\bibinfo {volume} {27}},\ \bibinfo {pages} {612}
  (\bibinfo {year} {2015})}\BibitemShut {NoStop}%
\bibitem [{\citenamefont {Huang}\ \emph {et~al.}(2017)\citenamefont {Huang},
  \citenamefont {Clark},\ and\ \citenamefont {Navarro-Moratalla}}]{Huang2017b}%
  \BibitemOpen
  \bibfield  {author} {\bibinfo {author} {\bibfnamefont {B.}~\bibnamefont
  {Huang}}, \bibinfo {author} {\bibfnamefont {G.}~\bibnamefont {Clark}}, \ and\
  \bibinfo {author} {\bibfnamefont {E.}~\bibnamefont {Navarro-Moratalla}},\
  }\href {\doibase 10.1038/nature22391} {\bibfield  {journal} {\bibinfo
  {journal} {Nature}\ }\textbf {\bibinfo {volume} {546}},\ \bibinfo {pages}
  {270} (\bibinfo {year} {2017})}\BibitemShut {NoStop}%
\bibitem [{\citenamefont {Webster}\ and\ \citenamefont {Yan}(2018)}]{RN15}%
  \BibitemOpen
  \bibfield  {author} {\bibinfo {author} {\bibfnamefont {L.}~\bibnamefont
  {Webster}}\ and\ \bibinfo {author} {\bibfnamefont {J.-A.}\ \bibnamefont
  {Yan}},\ }\href {\doibase 10.1103/PhysRevB.98.144411} {\bibfield  {journal}
  {\bibinfo  {journal} {Phys. Rev. B}\ }\textbf {\bibinfo {volume} {98}},\
  \bibinfo {pages} {144411} (\bibinfo {year} {2018})}\BibitemShut {NoStop}%
\bibitem [{\citenamefont {Hao}\ \emph {et~al.}(2018)\citenamefont {Hao},
  \citenamefont {Li},\ and\ \citenamefont {Zhang}}]{Hao2018}%
  \BibitemOpen
  \bibfield  {author} {\bibinfo {author} {\bibfnamefont {Z.}~\bibnamefont
  {Hao}}, \bibinfo {author} {\bibfnamefont {H.}~\bibnamefont {Li}}, \ and\
  \bibinfo {author} {\bibfnamefont {S.}~\bibnamefont {Zhang}},\ }\href
  {\doibase https://doi.org/10.1016/j.scib.2018.05.034} {\bibfield  {journal}
  {\bibinfo  {journal} {Sci. Bull.}\ }\textbf {\bibinfo {volume} {63}},\
  \bibinfo {pages} {825} (\bibinfo {year} {2018})}\BibitemShut {NoStop}%
\bibitem [{\citenamefont {Wang}\ \emph {et~al.}(2018)\citenamefont {Wang},
  \citenamefont {Zhang},\ and\ \citenamefont {Ding}}]{Wang2018}%
  \BibitemOpen
  \bibfield  {author} {\bibinfo {author} {\bibfnamefont {Z.}~\bibnamefont
  {Wang}}, \bibinfo {author} {\bibfnamefont {T.}~\bibnamefont {Zhang}}, \ and\
  \bibinfo {author} {\bibfnamefont {M.}~\bibnamefont {Ding}},\ }\href {\doibase
  https://doi.org/10.1038/s41565-018-0186-z} {\bibfield  {journal} {\bibinfo
  {journal} {Nat. Nanotechnol.}\ }\textbf {\bibinfo {volume} {13}},\ \bibinfo
  {pages} {554} (\bibinfo {year} {2018})}\BibitemShut {NoStop}%
\bibitem [{\citenamefont {Gong}\ \emph {et~al.}(2017)\citenamefont {Gong},
  \citenamefont {Li},\ and\ \citenamefont {Li}}]{Gong2017}%
  \BibitemOpen
  \bibfield  {author} {\bibinfo {author} {\bibfnamefont {C.}~\bibnamefont
  {Gong}}, \bibinfo {author} {\bibfnamefont {L.}~\bibnamefont {Li}}, \ and\
  \bibinfo {author} {\bibfnamefont {Z.}~\bibnamefont {Li}},\ }\href {\doibase
  https://doi.org/10.1038/s41565-018-0186-z} {\bibfield  {journal} {\bibinfo
  {journal} {Nat. Nanotechnol.}\ }\textbf {\bibinfo {volume} {546}},\ \bibinfo
  {pages} {554} (\bibinfo {year} {2017})}\BibitemShut {NoStop}%
\bibitem [{\citenamefont {Tu}\ \emph {et~al.}(2016)\citenamefont {Tu},
  \citenamefont {Hai},\ and\ \citenamefont {Anh}}]{Tu2016}%
  \BibitemOpen
  \bibfield  {author} {\bibinfo {author} {\bibfnamefont {N.}~\bibnamefont
  {Tu}}, \bibinfo {author} {\bibfnamefont {P.}~\bibnamefont {Hai}}, \ and\
  \bibinfo {author} {\bibfnamefont {L.}~\bibnamefont {Anh}},\ }\href {\doibase
  10.1063/1.4948692} {\bibfield  {journal} {\bibinfo  {journal} {Appl. Phys.
  Lett.}\ }\textbf {\bibinfo {volume} {108}},\ \bibinfo {pages} {192401}
  (\bibinfo {year} {2016})}\BibitemShut {NoStop}%
\bibitem [{\citenamefont {Furdyna}(1988)}]{Furdyna1988}%
  \BibitemOpen
  \bibfield  {author} {\bibinfo {author} {\bibfnamefont {J.}~\bibnamefont
  {Furdyna}},\ }\href {\doibase 10.1063/1.341700} {\bibfield  {journal}
  {\bibinfo  {journal} {J. Appl. Phys.}\ }\textbf {\bibinfo {volume} {64}},\
  \bibinfo {pages} {R29} (\bibinfo {year} {1988})}\BibitemShut {NoStop}%
\bibitem [{\citenamefont {Dalpian}\ \emph {et~al.}(2006)\citenamefont
  {Dalpian}, \citenamefont {Wei},\ and\ \citenamefont {Gong}}]{Dalpian2006}%
  \BibitemOpen
  \bibfield  {author} {\bibinfo {author} {\bibfnamefont {G.}~\bibnamefont
  {Dalpian}}, \bibinfo {author} {\bibfnamefont {S.-H.}\ \bibnamefont {Wei}}, \
  and\ \bibinfo {author} {\bibfnamefont {X.}~\bibnamefont {Gong}},\ }\href
  {\doibase https://doi.org/10.1016/j.ssc.2006.03.002} {\bibfield  {journal}
  {\bibinfo  {journal} {Solid State Commun.}\ }\textbf {\bibinfo {volume}
  {138}},\ \bibinfo {pages} {353} (\bibinfo {year} {2006})}\BibitemShut
  {NoStop}%
\bibitem [{\citenamefont {Pan}\ \emph {et~al.}(2008)\citenamefont {Pan},
  \citenamefont {Song},\ and\ \citenamefont {Liu}}]{Pan2008}%
  \BibitemOpen
  \bibfield  {author} {\bibinfo {author} {\bibfnamefont {F.}~\bibnamefont
  {Pan}}, \bibinfo {author} {\bibfnamefont {C.}~\bibnamefont {Song}}, \ and\
  \bibinfo {author} {\bibfnamefont {X.}~\bibnamefont {Liu}},\ }\href {\doibase
  https://doi.org/10.1016/j.mser.2008.04.002} {\bibfield  {journal} {\bibinfo
  {journal} {Mater. Sci. Eng. R-Rep.}\ }\textbf {\bibinfo {volume} {62}},\
  \bibinfo {pages} {1} (\bibinfo {year} {2008})}\BibitemShut {NoStop}%
\bibitem [{\citenamefont {Sato}\ \emph {et~al.}(2010)\citenamefont {Sato},
  \citenamefont {Bergqvist}, \citenamefont {Kudrnovsky}, \citenamefont
  {Dederichs}, \citenamefont {Eriksson}, \citenamefont {Turek}, \citenamefont
  {Sanyal}, \citenamefont {Bouzerar}, \citenamefont {Katayama-Yoshida},
  \citenamefont {Dinh}, \citenamefont {Fukushima}, \citenamefont {Kizaki},\
  and\ \citenamefont {Zeller}}]{Sato2010a}%
  \BibitemOpen
  \bibfield  {author} {\bibinfo {author} {\bibfnamefont {K.}~\bibnamefont
  {Sato}}, \bibinfo {author} {\bibfnamefont {L.}~\bibnamefont {Bergqvist}},
  \bibinfo {author} {\bibfnamefont {J.}~\bibnamefont {Kudrnovsky}}, \bibinfo
  {author} {\bibfnamefont {P.~H.}\ \bibnamefont {Dederichs}}, \bibinfo {author}
  {\bibfnamefont {O.}~\bibnamefont {Eriksson}}, \bibinfo {author}
  {\bibfnamefont {I.}~\bibnamefont {Turek}}, \bibinfo {author} {\bibfnamefont
  {B.}~\bibnamefont {Sanyal}}, \bibinfo {author} {\bibfnamefont
  {G.}~\bibnamefont {Bouzerar}}, \bibinfo {author} {\bibfnamefont
  {H.}~\bibnamefont {Katayama-Yoshida}}, \bibinfo {author} {\bibfnamefont
  {V.~A.}\ \bibnamefont {Dinh}}, \bibinfo {author} {\bibfnamefont
  {T.}~\bibnamefont {Fukushima}}, \bibinfo {author} {\bibfnamefont
  {H.}~\bibnamefont {Kizaki}}, \ and\ \bibinfo {author} {\bibfnamefont
  {R.}~\bibnamefont {Zeller}},\ }\href {\doibase 10.1103/RevModPhys.82.1633}
  {\bibfield  {journal} {\bibinfo  {journal} {Rev. Mod. Phys.}\ }\textbf
  {\bibinfo {volume} {82}},\ \bibinfo {pages} {1633} (\bibinfo {year}
  {2010})}\BibitemShut {NoStop}%
\bibitem [{\citenamefont {Dietl}(2010)}]{Dietl2010}%
  \BibitemOpen
  \bibfield  {author} {\bibinfo {author} {\bibfnamefont {T.}~\bibnamefont
  {Dietl}},\ }\href {\doibase https://doi.org/10.1038/nmat2898} {\bibfield
  {journal} {\bibinfo  {journal} {Nat. Mater.}\ }\textbf {\bibinfo {volume}
  {9}},\ \bibinfo {pages} {965} (\bibinfo {year} {2010})}\BibitemShut {NoStop}%
\bibitem [{\citenamefont {Wei}\ and\ \citenamefont {Dalpian}(2008)}]{Wei2008}%
  \BibitemOpen
  \bibfield  {author} {\bibinfo {author} {\bibfnamefont {S.-H.}\ \bibnamefont
  {Wei}}\ and\ \bibinfo {author} {\bibfnamefont {G.~M.}\ \bibnamefont
  {Dalpian}},\ }in\ \href {\doibase https://doi.org/10.1117/12.763494} {\emph
  {\bibinfo {booktitle} {Proceedings of the integrated optoelectronic
  devices}}}\ (\bibinfo  {publisher} {SPIE},\ \bibinfo {year} {2008})\ pp.\
  \bibinfo {pages} {85--95}\BibitemShut {NoStop}%
\bibitem [{\citenamefont {Bouzerar}\ \emph {et~al.}(2007)\citenamefont
  {Bouzerar}, \citenamefont {Bouzerar},\ and\ \citenamefont
  {Ziman}}]{Bouzerar2007}%
  \BibitemOpen
  \bibfield  {author} {\bibinfo {author} {\bibfnamefont {R.}~\bibnamefont
  {Bouzerar}}, \bibinfo {author} {\bibfnamefont {G.}~\bibnamefont {Bouzerar}},
  \ and\ \bibinfo {author} {\bibfnamefont {T.}~\bibnamefont {Ziman}},\ }\href
  {\doibase http://dx.doi.org/10.1209/0295-5075/78/67003} {\bibfield  {journal}
  {\bibinfo  {journal} {Europhys. Lett.}\ }\textbf {\bibinfo {volume} {78}},\
  \bibinfo {pages} {67003} (\bibinfo {year} {2007})}\BibitemShut {NoStop}%
\bibitem [{\citenamefont {Chen}\ \emph {et~al.}(2011)\citenamefont {Chen},
  \citenamefont {Yang},\ and\ \citenamefont {Yang}}]{Chen2011}%
  \BibitemOpen
  \bibfield  {author} {\bibinfo {author} {\bibfnamefont {L.}~\bibnamefont
  {Chen}}, \bibinfo {author} {\bibfnamefont {X.}~\bibnamefont {Yang}}, \ and\
  \bibinfo {author} {\bibfnamefont {F.}~\bibnamefont {Yang}},\ }\href {\doibase
  http://dx.doi.org/10.1021/nl201187m} {\bibfield  {journal} {\bibinfo
  {journal} {Nano Lett.}\ }\textbf {\bibinfo {volume} {11}},\ \bibinfo {pages}
  {2584} (\bibinfo {year} {2011})}\BibitemShut {NoStop}%
\bibitem [{\citenamefont {Zutic}\ and\ \citenamefont {Zhou}(2018)}]{Zutic2018}%
  \BibitemOpen
  \bibfield  {author} {\bibinfo {author} {\bibfnamefont {I.}~\bibnamefont
  {Zutic}}\ and\ \bibinfo {author} {\bibfnamefont {T.}~\bibnamefont {Zhou}},\
  }\href {\doibase https://doi.org/10.1007/s11433-018-9191-0} {\bibfield
  {journal} {\bibinfo  {journal} {Sci. China Phys. Mech. Astron.}\ }\textbf
  {\bibinfo {volume} {61}},\ \bibinfo {pages} {67031} (\bibinfo {year}
  {2018})}\BibitemShut {NoStop}%
\bibitem [{\citenamefont {Bonilla}\ \emph {et~al.}(2018)\citenamefont
  {Bonilla}, \citenamefont {Kolekar},\ and\ \citenamefont {Ma}}]{Bonilla2018}%
  \BibitemOpen
  \bibfield  {author} {\bibinfo {author} {\bibfnamefont {M.}~\bibnamefont
  {Bonilla}}, \bibinfo {author} {\bibfnamefont {S.}~\bibnamefont {Kolekar}}, \
  and\ \bibinfo {author} {\bibfnamefont {Y.}~\bibnamefont {Ma}},\ }\href
  {\doibase https://doi.org/10.1038/s41565-018-0063-9} {\bibfield  {journal}
  {\bibinfo  {journal} {Nat. Nanotechnol.}\ }\textbf {\bibinfo {volume} {13}},\
  \bibinfo {pages} {289} (\bibinfo {year} {2018})}\BibitemShut {NoStop}%
\bibitem [{\citenamefont {Liu}\ \emph {et~al.}(2018{\natexlab{a}})\citenamefont
  {Liu}, \citenamefont {Wu},\ and\ \citenamefont {Shao}}]{Liu2018b}%
  \BibitemOpen
  \bibfield  {author} {\bibinfo {author} {\bibfnamefont {Z.}~\bibnamefont
  {Liu}}, \bibinfo {author} {\bibfnamefont {X.}~\bibnamefont {Wu}}, \ and\
  \bibinfo {author} {\bibfnamefont {Y.}~\bibnamefont {Shao}},\ }\href {\doibase
  https://doi.org/10.1016/j.scib.2018.03.008} {\bibfield  {journal} {\bibinfo
  {journal} {Sci. Bull.}\ }\textbf {\bibinfo {volume} {63}},\ \bibinfo {pages}
  {419} (\bibinfo {year} {2018}{\natexlab{a}})}\BibitemShut {NoStop}%
\bibitem [{\citenamefont {Tan}\ \emph {et~al.}(2018)\citenamefont {Tan},
  \citenamefont {Lee},\ and\ \citenamefont {Jung}}]{Tan2018}%
  \BibitemOpen
  \bibfield  {author} {\bibinfo {author} {\bibfnamefont {C.}~\bibnamefont
  {Tan}}, \bibinfo {author} {\bibfnamefont {J.}~\bibnamefont {Lee}}, \ and\
  \bibinfo {author} {\bibfnamefont {S.-G.}\ \bibnamefont {Jung}},\ }\href
  {\doibase https://doi.org/10.1038/s41467-018-04018-w} {\bibfield  {journal}
  {\bibinfo  {journal} {Nat. Commun.}\ }\textbf {\bibinfo {volume} {9}},\
  \bibinfo {pages} {1554} (\bibinfo {year} {2018})}\BibitemShut {NoStop}%
\bibitem [{\citenamefont {Deng}\ \emph {et~al.}(2018)\citenamefont {Deng},
  \citenamefont {Yu},\ and\ \citenamefont {Song}}]{Deng2018}%
  \BibitemOpen
  \bibfield  {author} {\bibinfo {author} {\bibfnamefont {Y.}~\bibnamefont
  {Deng}}, \bibinfo {author} {\bibfnamefont {Y.}~\bibnamefont {Yu}}, \ and\
  \bibinfo {author} {\bibfnamefont {Y.}~\bibnamefont {Song}},\ }\href {\doibase
  https://doi.org/10.1038/s41586-018-0626-9} {\bibfield  {journal} {\bibinfo
  {journal} {Nature}\ }\textbf {\bibinfo {volume} {563}},\ \bibinfo {pages}
  {94} (\bibinfo {year} {2018})}\BibitemShut {NoStop}%
\bibitem [{\citenamefont {Fei}\ \emph {et~al.}(2018)\citenamefont {Fei},
  \citenamefont {Huang},\ and\ \citenamefont {Malinowski}}]{Fei2018}%
  \BibitemOpen
  \bibfield  {author} {\bibinfo {author} {\bibfnamefont {Z.}~\bibnamefont
  {Fei}}, \bibinfo {author} {\bibfnamefont {B.}~\bibnamefont {Huang}}, \ and\
  \bibinfo {author} {\bibfnamefont {P.}~\bibnamefont {Malinowski}},\ }\href
  {\doibase https://doi.org/10.1038/s41563-018-0149-7} {\bibfield  {journal}
  {\bibinfo  {journal} {Nat. Mater.}\ }\textbf {\bibinfo {volume} {17}},\
  \bibinfo {pages} {778} (\bibinfo {year} {2018})}\BibitemShut {NoStop}%
\bibitem [{\citenamefont {Wang}\ \emph {et~al.}(2019)\citenamefont {Wang},
  \citenamefont {Zhou}, \citenamefont {Zhou}, \citenamefont {Tong},
  \citenamefont {Lu},\ and\ \citenamefont {Ji}}]{Wang2019d}%
  \BibitemOpen
  \bibfield  {author} {\bibinfo {author} {\bibfnamefont {C.}~\bibnamefont
  {Wang}}, \bibinfo {author} {\bibfnamefont {X.}~\bibnamefont {Zhou}}, \bibinfo
  {author} {\bibfnamefont {L.}~\bibnamefont {Zhou}}, \bibinfo {author}
  {\bibfnamefont {N.-H.}\ \bibnamefont {Tong}}, \bibinfo {author}
  {\bibfnamefont {Z.-Y.}\ \bibnamefont {Lu}}, \ and\ \bibinfo {author}
  {\bibfnamefont {W.}~\bibnamefont {Ji}},\ }\href {\doibase
  https://doi.org/10.1016/j.scib.2019.02.011} {\bibfield  {journal} {\bibinfo
  {journal} {Sci. Bull.}\ }\textbf {\bibinfo {volume} {64}},\ \bibinfo {pages}
  {293} (\bibinfo {year} {2019})}\BibitemShut {NoStop}%
\bibitem [{\citenamefont {Göbel}\ \emph {et~al.}(2021)\citenamefont {Göbel},
  \citenamefont {Mertig},\ and\ \citenamefont {Tretiakov}}]{RN2009}%
  \BibitemOpen
  \bibfield  {author} {\bibinfo {author} {\bibfnamefont {B.}~\bibnamefont
  {Göbel}}, \bibinfo {author} {\bibfnamefont {I.}~\bibnamefont {Mertig}}, \
  and\ \bibinfo {author} {\bibfnamefont {O.~A.}\ \bibnamefont {Tretiakov}},\
  }\href {\doibase https://doi.org/10.1016/j.physrep.2020.10.001} {\bibfield
  {journal} {\bibinfo  {journal} {Phys. Rep.}\ }\textbf {\bibinfo {volume}
  {895}},\ \bibinfo {pages} {1} (\bibinfo {year} {2021})}\BibitemShut {NoStop}%
\bibitem [{\citenamefont {Liu}\ \emph {et~al.}(2020)\citenamefont {Liu},
  \citenamefont {Qian},\ and\ \citenamefont {Zhu}}]{RN120}%
  \BibitemOpen
  \bibfield  {author} {\bibinfo {author} {\bibfnamefont {Y.}~\bibnamefont
  {Liu}}, \bibinfo {author} {\bibfnamefont {Z.~H.}\ \bibnamefont {Qian}}, \
  and\ \bibinfo {author} {\bibfnamefont {J.~G.}\ \bibnamefont {Zhu}},\ }\href
  {\doibase http://dx.doi.org/10.7498/aps.69.20200984} {\bibfield  {journal}
  {\bibinfo  {journal} {Acta Phys. Sin.}\ }\textbf {\bibinfo {volume} {69}},\
  \bibinfo {pages} {231201} (\bibinfo {year} {2020})}\BibitemShut {NoStop}%
\bibitem [{\citenamefont {Liu}\ and\ \citenamefont {Zang}(2018)}]{RN121}%
  \BibitemOpen
  \bibfield  {author} {\bibinfo {author} {\bibfnamefont {Y.-Z.}\ \bibnamefont
  {Liu}}\ and\ \bibinfo {author} {\bibfnamefont {J.}~\bibnamefont {Zang}},\
  }\href {\doibase https://doi.org/10.7498/aps.67.20180619} {\bibfield
  {journal} {\bibinfo  {journal} {Acta Phys. Sin.}\ }\textbf {\bibinfo {volume}
  {67}},\ \bibinfo {pages} {131201} (\bibinfo {year} {2018})}\BibitemShut
  {NoStop}%
\bibitem [{\citenamefont {Hoffmann}\ \emph {et~al.}(2017)\citenamefont
  {Hoffmann}, \citenamefont {Zimmermann}, \citenamefont {Muller}, \citenamefont
  {Schurhoff}, \citenamefont {Kiselev}, \citenamefont {Melcher},\ and\
  \citenamefont {Blugel}}]{RN109}%
  \BibitemOpen
  \bibfield  {author} {\bibinfo {author} {\bibfnamefont {M.}~\bibnamefont
  {Hoffmann}}, \bibinfo {author} {\bibfnamefont {B.}~\bibnamefont
  {Zimmermann}}, \bibinfo {author} {\bibfnamefont {G.~P.}\ \bibnamefont
  {Muller}}, \bibinfo {author} {\bibfnamefont {D.}~\bibnamefont {Schurhoff}},
  \bibinfo {author} {\bibfnamefont {N.~S.}\ \bibnamefont {Kiselev}}, \bibinfo
  {author} {\bibfnamefont {C.}~\bibnamefont {Melcher}}, \ and\ \bibinfo
  {author} {\bibfnamefont {S.}~\bibnamefont {Blugel}},\ }\href {\doibase
  https://doi.org/10.1038/s41467-017-00313-0} {\bibfield  {journal} {\bibinfo
  {journal} {Nat. Commun.}\ }\textbf {\bibinfo {volume} {8}},\ \bibinfo {pages}
  {308} (\bibinfo {year} {2017})}\BibitemShut {NoStop}%
\bibitem [{\citenamefont {Mochizuki}\ and\ \citenamefont {Seki}(2013)}]{RN887}%
  \BibitemOpen
  \bibfield  {author} {\bibinfo {author} {\bibfnamefont {M.}~\bibnamefont
  {Mochizuki}}\ and\ \bibinfo {author} {\bibfnamefont {S.}~\bibnamefont
  {Seki}},\ }\href {\doibase https://doi.org/10.1103/PhysRevB.87.134403}
  {\bibfield  {journal} {\bibinfo  {journal} {Phys. Rev. B}\ }\textbf {\bibinfo
  {volume} {87}},\ \bibinfo {pages} {134403} (\bibinfo {year}
  {2013})}\BibitemShut {NoStop}%
\bibitem [{\citenamefont {Hou}\ \emph {et~al.}(2018)\citenamefont {Hou},
  \citenamefont {Ding}, \citenamefont {Li}, \citenamefont {Xu}, \citenamefont
  {Wang},\ and\ \citenamefont {Wu}}]{RN1986}%
  \BibitemOpen
  \bibfield  {author} {\bibinfo {author} {\bibfnamefont {Z.-P.}\ \bibnamefont
  {Hou}}, \bibinfo {author} {\bibfnamefont {B.}~\bibnamefont {Ding}}, \bibinfo
  {author} {\bibfnamefont {H.}~\bibnamefont {Li}}, \bibinfo {author}
  {\bibfnamefont {G.-Z.}\ \bibnamefont {Xu}}, \bibinfo {author} {\bibfnamefont
  {W.-H.}\ \bibnamefont {Wang}}, \ and\ \bibinfo {author} {\bibfnamefont
  {G.-H.}\ \bibnamefont {Wu}},\ }\href {\doibase
  https://doi.org/10.7498/aps.67.20180419} {\bibfield  {journal} {\bibinfo
  {journal} {Acta Phys. Sin.}\ }\textbf {\bibinfo {volume} {67}},\ \bibinfo
  {pages} {137509} (\bibinfo {year} {2018})}\BibitemShut {NoStop}%
\bibitem [{\citenamefont {Parkin}\ \emph {et~al.}(2008)\citenamefont {Parkin},
  \citenamefont {Hayashi},\ and\ \citenamefont {Thomas}}]{RN1988}%
  \BibitemOpen
  \bibfield  {author} {\bibinfo {author} {\bibfnamefont {S.~S.}\ \bibnamefont
  {Parkin}}, \bibinfo {author} {\bibfnamefont {M.}~\bibnamefont {Hayashi}}, \
  and\ \bibinfo {author} {\bibfnamefont {L.}~\bibnamefont {Thomas}},\ }\href
  {\doibase https://doi.org/10.1126/science.1145799} {\bibfield  {journal}
  {\bibinfo  {journal} {Science}\ }\textbf {\bibinfo {volume} {320}},\ \bibinfo
  {pages} {190} (\bibinfo {year} {2008})}\BibitemShut {NoStop}%
\bibitem [{\citenamefont {Zhang}\ \emph
  {et~al.}(2020{\natexlab{a}})\citenamefont {Zhang}, \citenamefont {Zhou},
  \citenamefont {Mee~Song}, \citenamefont {Park}, \citenamefont {Xia},
  \citenamefont {Ezawa}, \citenamefont {Liu}, \citenamefont {Zhao},
  \citenamefont {Zhao},\ and\ \citenamefont {Woo}}]{RN116}%
  \BibitemOpen
  \bibfield  {author} {\bibinfo {author} {\bibfnamefont {X.}~\bibnamefont
  {Zhang}}, \bibinfo {author} {\bibfnamefont {Y.}~\bibnamefont {Zhou}},
  \bibinfo {author} {\bibfnamefont {K.}~\bibnamefont {Mee~Song}}, \bibinfo
  {author} {\bibfnamefont {T.~E.}\ \bibnamefont {Park}}, \bibinfo {author}
  {\bibfnamefont {J.}~\bibnamefont {Xia}}, \bibinfo {author} {\bibfnamefont
  {M.}~\bibnamefont {Ezawa}}, \bibinfo {author} {\bibfnamefont
  {X.}~\bibnamefont {Liu}}, \bibinfo {author} {\bibfnamefont {W.}~\bibnamefont
  {Zhao}}, \bibinfo {author} {\bibfnamefont {G.}~\bibnamefont {Zhao}}, \ and\
  \bibinfo {author} {\bibfnamefont {S.}~\bibnamefont {Woo}},\ }\href {\doibase
  https://doi.org/10.1088/1361-648x/ab5488} {\bibfield  {journal} {\bibinfo
  {journal} {J. Phys.: Condens. Matter}\ }\textbf {\bibinfo {volume} {32}},\
  \bibinfo {pages} {143001} (\bibinfo {year} {2020}{\natexlab{a}})}\BibitemShut
  {NoStop}%
\bibitem [{\citenamefont {Wang}\ \emph {et~al.}(2021)\citenamefont {Wang},
  \citenamefont {Strungaru}, \citenamefont {Ruta}, \citenamefont {Meo},
  \citenamefont {Zhou}, \citenamefont {Deak}, \citenamefont {Szunyogh},
  \citenamefont {Gavriloaea}, \citenamefont {Moreno}, \citenamefont
  {Chubykalo-Fesenko}, \citenamefont {Wu}, \citenamefont {Xu}, \citenamefont
  {Evans},\ and\ \citenamefont {Chantrell}}]{RN3121}%
  \BibitemOpen
  \bibfield  {author} {\bibinfo {author} {\bibfnamefont {J.}~\bibnamefont
  {Wang}}, \bibinfo {author} {\bibfnamefont {M.}~\bibnamefont {Strungaru}},
  \bibinfo {author} {\bibfnamefont {S.}~\bibnamefont {Ruta}}, \bibinfo {author}
  {\bibfnamefont {A.}~\bibnamefont {Meo}}, \bibinfo {author} {\bibfnamefont
  {Y.}~\bibnamefont {Zhou}}, \bibinfo {author} {\bibfnamefont {A.}~\bibnamefont
  {Deak}}, \bibinfo {author} {\bibfnamefont {L.}~\bibnamefont {Szunyogh}},
  \bibinfo {author} {\bibfnamefont {P.-I.}\ \bibnamefont {Gavriloaea}},
  \bibinfo {author} {\bibfnamefont {R.}~\bibnamefont {Moreno}}, \bibinfo
  {author} {\bibfnamefont {O.}~\bibnamefont {Chubykalo-Fesenko}}, \bibinfo
  {author} {\bibfnamefont {J.}~\bibnamefont {Wu}}, \bibinfo {author}
  {\bibfnamefont {Y.}~\bibnamefont {Xu}}, \bibinfo {author} {\bibfnamefont
  {R.~F.~L.}\ \bibnamefont {Evans}}, \ and\ \bibinfo {author} {\bibfnamefont
  {R.~W.}\ \bibnamefont {Chantrell}},\ }\href {\doibase
  https://doi.org/10.1103/PhysRevB.104.054420} {\bibfield  {journal} {\bibinfo
  {journal} {Phys. Rev. B}\ }\textbf {\bibinfo {volume} {104}},\ \bibinfo
  {pages} {054420} (\bibinfo {year} {2021})}\BibitemShut {NoStop}%
\bibitem [{\citenamefont {Wu}\ \emph {et~al.}(2021)\citenamefont {Wu},
  \citenamefont {Wen}, \citenamefont {Chen},\ and\ \citenamefont
  {Zheng}}]{RN3542}%
  \BibitemOpen
  \bibfield  {author} {\bibinfo {author} {\bibfnamefont {Y.}~\bibnamefont
  {Wu}}, \bibinfo {author} {\bibfnamefont {H.}~\bibnamefont {Wen}}, \bibinfo
  {author} {\bibfnamefont {W.}~\bibnamefont {Chen}}, \ and\ \bibinfo {author}
  {\bibfnamefont {Y.}~\bibnamefont {Zheng}},\ }\href {\doibase
  https://doi.org/10.1103/PhysRevLett.127.097201} {\bibfield  {journal}
  {\bibinfo  {journal} {Phys. Rev. Lett.}\ }\textbf {\bibinfo {volume} {127}},\
  \bibinfo {pages} {097201} (\bibinfo {year} {2021})}\BibitemShut {NoStop}%
\bibitem [{\citenamefont {Fert}\ \emph {et~al.}(2013)\citenamefont {Fert},
  \citenamefont {Cros},\ and\ \citenamefont {Sampaio}}]{RN1040}%
  \BibitemOpen
  \bibfield  {author} {\bibinfo {author} {\bibfnamefont {A.}~\bibnamefont
  {Fert}}, \bibinfo {author} {\bibfnamefont {V.}~\bibnamefont {Cros}}, \ and\
  \bibinfo {author} {\bibfnamefont {J.}~\bibnamefont {Sampaio}},\ }\href
  {\doibase https://doi.org/10.1038/nnano.2013.29} {\bibfield  {journal}
  {\bibinfo  {journal} {Nat. Nanotechnol.}\ }\textbf {\bibinfo {volume} {8}},\
  \bibinfo {pages} {152} (\bibinfo {year} {2013})}\BibitemShut {NoStop}%
\bibitem [{\citenamefont {Yu}\ \emph {et~al.}(2021)\citenamefont {Yu},
  \citenamefont {Kagawa}, \citenamefont {Seki}, \citenamefont {Kubota},
  \citenamefont {Masell}, \citenamefont {Yasin}, \citenamefont {Nakajima},
  \citenamefont {Nakamura}, \citenamefont {Kawasaki}, \citenamefont {Nagaosa},\
  and\ \citenamefont {Tokura}}]{RN3177}%
  \BibitemOpen
  \bibfield  {author} {\bibinfo {author} {\bibfnamefont {X.}~\bibnamefont
  {Yu}}, \bibinfo {author} {\bibfnamefont {F.}~\bibnamefont {Kagawa}}, \bibinfo
  {author} {\bibfnamefont {S.}~\bibnamefont {Seki}}, \bibinfo {author}
  {\bibfnamefont {M.}~\bibnamefont {Kubota}}, \bibinfo {author} {\bibfnamefont
  {J.}~\bibnamefont {Masell}}, \bibinfo {author} {\bibfnamefont {F.~S.}\
  \bibnamefont {Yasin}}, \bibinfo {author} {\bibfnamefont {K.}~\bibnamefont
  {Nakajima}}, \bibinfo {author} {\bibfnamefont {M.}~\bibnamefont {Nakamura}},
  \bibinfo {author} {\bibfnamefont {M.}~\bibnamefont {Kawasaki}}, \bibinfo
  {author} {\bibfnamefont {N.}~\bibnamefont {Nagaosa}}, \ and\ \bibinfo
  {author} {\bibfnamefont {Y.}~\bibnamefont {Tokura}},\ }\href {\doibase
  10.1038/s41467-021-25291-2} {\bibfield  {journal} {\bibinfo  {journal} {Nat.
  Commun.}\ }\textbf {\bibinfo {volume} {12}},\ \bibinfo {pages} {5079}
  (\bibinfo {year} {2021})}\BibitemShut {NoStop}%
\bibitem [{\citenamefont {I.}(1958)}]{RN130}%
  \BibitemOpen
  \bibfield  {author} {\bibinfo {author} {\bibfnamefont {D.}~\bibnamefont
  {I.}},\ }\href {\doibase 10.1016/0927-0256(96)00008-0} {\bibfield  {journal}
  {\bibinfo  {journal} {J. Phys. Chem. Solids}\ }\textbf {\bibinfo {volume}
  {4}},\ \bibinfo {pages} {241} (\bibinfo {year} {1958})}\BibitemShut {NoStop}%
\bibitem [{\citenamefont {Moriya}(1960)}]{RN131}%
  \BibitemOpen
  \bibfield  {author} {\bibinfo {author} {\bibfnamefont {T.}~\bibnamefont
  {Moriya}},\ }\href {\doibase 10.1103/PhysRev.120.91} {\bibfield  {journal}
  {\bibinfo  {journal} {Phys. Rev.}\ }\textbf {\bibinfo {volume} {120}},\
  \bibinfo {pages} {91} (\bibinfo {year} {1960})}\BibitemShut {NoStop}%
\bibitem [{\citenamefont {Nagaosa}\ and\ \citenamefont
  {Tokura}(2013)}]{Nagaosa2013}%
  \BibitemOpen
  \bibfield  {author} {\bibinfo {author} {\bibfnamefont {N.}~\bibnamefont
  {Nagaosa}}\ and\ \bibinfo {author} {\bibfnamefont {Y.}~\bibnamefont
  {Tokura}},\ }\href {\doibase https://doi.org/10.1038/nnano.2013.243}
  {\bibfield  {journal} {\bibinfo  {journal} {Nat. Nanotechnol.}\ }\textbf
  {\bibinfo {volume} {8}},\ \bibinfo {pages} {899} (\bibinfo {year}
  {2013})}\BibitemShut {NoStop}%
\bibitem [{\citenamefont {Belabbes}\ \emph {et~al.}(2016)\citenamefont
  {Belabbes}, \citenamefont {Bihlmayer}, \citenamefont {Bechstedt},
  \citenamefont {Blugel},\ and\ \citenamefont {Manchon}}]{RN24}%
  \BibitemOpen
  \bibfield  {author} {\bibinfo {author} {\bibfnamefont {A.}~\bibnamefont
  {Belabbes}}, \bibinfo {author} {\bibfnamefont {G.}~\bibnamefont {Bihlmayer}},
  \bibinfo {author} {\bibfnamefont {F.}~\bibnamefont {Bechstedt}}, \bibinfo
  {author} {\bibfnamefont {S.}~\bibnamefont {Blugel}}, \ and\ \bibinfo {author}
  {\bibfnamefont {A.}~\bibnamefont {Manchon}},\ }\href {\doibase
  10.1103/PhysRevLett.117.247202} {\bibfield  {journal} {\bibinfo  {journal}
  {Phys. Rev. Lett.}\ }\textbf {\bibinfo {volume} {117}},\ \bibinfo {pages}
  {247202} (\bibinfo {year} {2016})}\BibitemShut {NoStop}%
\bibitem [{\citenamefont {Yang}\ \emph {et~al.}(2015)\citenamefont {Yang},
  \citenamefont {Thiaville}, \citenamefont {Rohart}, \citenamefont {Fert},\
  and\ \citenamefont {Chshiev}}]{RN33}%
  \BibitemOpen
  \bibfield  {author} {\bibinfo {author} {\bibfnamefont {H.}~\bibnamefont
  {Yang}}, \bibinfo {author} {\bibfnamefont {A.}~\bibnamefont {Thiaville}},
  \bibinfo {author} {\bibfnamefont {S.}~\bibnamefont {Rohart}}, \bibinfo
  {author} {\bibfnamefont {A.}~\bibnamefont {Fert}}, \ and\ \bibinfo {author}
  {\bibfnamefont {M.}~\bibnamefont {Chshiev}},\ }\href {\doibase
  10.1103/PhysRevLett.115.267210} {\bibfield  {journal} {\bibinfo  {journal}
  {Phys. Rev. Lett.}\ }\textbf {\bibinfo {volume} {115}},\ \bibinfo {pages}
  {267210} (\bibinfo {year} {2015})}\BibitemShut {NoStop}%
\bibitem [{\citenamefont {Yang}\ \emph {et~al.}(2018)\citenamefont {Yang},
  \citenamefont {Chen}, \citenamefont {Cotta}, \citenamefont {N’Diaye},
  \citenamefont {Nikolaev}, \citenamefont {Soares}, \citenamefont {Macedo},
  \citenamefont {Liu}, \citenamefont {Schmid}, \citenamefont {Fert},\ and\
  \citenamefont {Chshiev}}]{RN32}%
  \BibitemOpen
  \bibfield  {author} {\bibinfo {author} {\bibfnamefont {H.}~\bibnamefont
  {Yang}}, \bibinfo {author} {\bibfnamefont {G.}~\bibnamefont {Chen}}, \bibinfo
  {author} {\bibfnamefont {A.~A.~C.}\ \bibnamefont {Cotta}}, \bibinfo {author}
  {\bibfnamefont {A.~T.}\ \bibnamefont {N’Diaye}}, \bibinfo {author}
  {\bibfnamefont {S.~A.}\ \bibnamefont {Nikolaev}}, \bibinfo {author}
  {\bibfnamefont {E.~A.}\ \bibnamefont {Soares}}, \bibinfo {author}
  {\bibfnamefont {W.~A.~A.}\ \bibnamefont {Macedo}}, \bibinfo {author}
  {\bibfnamefont {K.}~\bibnamefont {Liu}}, \bibinfo {author} {\bibfnamefont
  {A.~K.}\ \bibnamefont {Schmid}}, \bibinfo {author} {\bibfnamefont
  {A.}~\bibnamefont {Fert}}, \ and\ \bibinfo {author} {\bibfnamefont
  {M.}~\bibnamefont {Chshiev}},\ }\href {\doibase 10.1038/s41563-018-0079-4}
  {\bibfield  {journal} {\bibinfo  {journal} {Nat. Mater.}\ }\textbf {\bibinfo
  {volume} {17}},\ \bibinfo {pages} {605} (\bibinfo {year} {2018})}\BibitemShut
  {NoStop}%
\bibitem [{\citenamefont {Morshed}\ \emph {et~al.}(2021)\citenamefont
  {Morshed}, \citenamefont {Khoo}, \citenamefont {Quessab}, \citenamefont {Xu},
  \citenamefont {Laskowski}, \citenamefont {Balachandran}, \citenamefont
  {Kent},\ and\ \citenamefont {Ghosh}}]{RN2001}%
  \BibitemOpen
  \bibfield  {author} {\bibinfo {author} {\bibfnamefont {M.~G.}\ \bibnamefont
  {Morshed}}, \bibinfo {author} {\bibfnamefont {K.~H.}\ \bibnamefont {Khoo}},
  \bibinfo {author} {\bibfnamefont {Y.}~\bibnamefont {Quessab}}, \bibinfo
  {author} {\bibfnamefont {J.-W.}\ \bibnamefont {Xu}}, \bibinfo {author}
  {\bibfnamefont {R.}~\bibnamefont {Laskowski}}, \bibinfo {author}
  {\bibfnamefont {P.~V.}\ \bibnamefont {Balachandran}}, \bibinfo {author}
  {\bibfnamefont {A.~D.}\ \bibnamefont {Kent}}, \ and\ \bibinfo {author}
  {\bibfnamefont {A.~W.}\ \bibnamefont {Ghosh}},\ }\href {\doibase
  10.1103/PhysRevB.103.174414} {\bibfield  {journal} {\bibinfo  {journal}
  {Phys. Rev. B}\ }\textbf {\bibinfo {volume} {103}},\ \bibinfo {pages}
  {174414} (\bibinfo {year} {2021})}\BibitemShut {NoStop}%
\bibitem [{\citenamefont {Liang}\ \emph {et~al.}(2020)\citenamefont {Liang},
  \citenamefont {Wang}, \citenamefont {Du}, \citenamefont {Hallal},
  \citenamefont {Garcia}, \citenamefont {Chshiev}, \citenamefont {Fert},\ and\
  \citenamefont {Yang}}]{RN30}%
  \BibitemOpen
  \bibfield  {author} {\bibinfo {author} {\bibfnamefont {J.}~\bibnamefont
  {Liang}}, \bibinfo {author} {\bibfnamefont {W.}~\bibnamefont {Wang}},
  \bibinfo {author} {\bibfnamefont {H.}~\bibnamefont {Du}}, \bibinfo {author}
  {\bibfnamefont {A.}~\bibnamefont {Hallal}}, \bibinfo {author} {\bibfnamefont
  {K.}~\bibnamefont {Garcia}}, \bibinfo {author} {\bibfnamefont
  {M.}~\bibnamefont {Chshiev}}, \bibinfo {author} {\bibfnamefont
  {A.}~\bibnamefont {Fert}}, \ and\ \bibinfo {author} {\bibfnamefont
  {H.}~\bibnamefont {Yang}},\ }\href {\doibase 10.1103/PhysRevB.101.184401}
  {\bibfield  {journal} {\bibinfo  {journal} {Phys. Rev. B}\ }\textbf {\bibinfo
  {volume} {101}},\ \bibinfo {pages} {184401} (\bibinfo {year}
  {2020})}\BibitemShut {NoStop}%
\bibitem [{\citenamefont {Cui}\ \emph {et~al.}(2020)\citenamefont {Cui},
  \citenamefont {Liang}, \citenamefont {Shao}, \citenamefont {Cui},\ and\
  \citenamefont {Yang}}]{RN31}%
  \BibitemOpen
  \bibfield  {author} {\bibinfo {author} {\bibfnamefont {Q.}~\bibnamefont
  {Cui}}, \bibinfo {author} {\bibfnamefont {J.}~\bibnamefont {Liang}}, \bibinfo
  {author} {\bibfnamefont {Z.}~\bibnamefont {Shao}}, \bibinfo {author}
  {\bibfnamefont {P.}~\bibnamefont {Cui}}, \ and\ \bibinfo {author}
  {\bibfnamefont {H.}~\bibnamefont {Yang}},\ }\href {\doibase
  10.1103/PhysRevB.102.094425} {\bibfield  {journal} {\bibinfo  {journal}
  {Phys. Rev. B}\ }\textbf {\bibinfo {volume} {102}},\ \bibinfo {pages}
  {094425} (\bibinfo {year} {2020})}\BibitemShut {NoStop}%
\bibitem [{\citenamefont {Xu}\ \emph {et~al.}(2020)\citenamefont {Xu},
  \citenamefont {Feng}, \citenamefont {Prokhorenko}, \citenamefont {Nahas},
  \citenamefont {Xiang},\ and\ \citenamefont {Bellaiche}}]{RN1041}%
  \BibitemOpen
  \bibfield  {author} {\bibinfo {author} {\bibfnamefont {C.}~\bibnamefont
  {Xu}}, \bibinfo {author} {\bibfnamefont {J.}~\bibnamefont {Feng}}, \bibinfo
  {author} {\bibfnamefont {S.}~\bibnamefont {Prokhorenko}}, \bibinfo {author}
  {\bibfnamefont {Y.}~\bibnamefont {Nahas}}, \bibinfo {author} {\bibfnamefont
  {H.}~\bibnamefont {Xiang}}, \ and\ \bibinfo {author} {\bibfnamefont
  {L.}~\bibnamefont {Bellaiche}},\ }\href {\doibase
  10.1103/PhysRevB.101.060404} {\bibfield  {journal} {\bibinfo  {journal}
  {Phys. Rev. B}\ }\textbf {\bibinfo {volume} {101}},\ \bibinfo {pages}
  {060404(R)} (\bibinfo {year} {2020})}\BibitemShut {NoStop}%
\bibitem [{\citenamefont {Zhang}\ \emph
  {et~al.}(2020{\natexlab{b}})\citenamefont {Zhang}, \citenamefont {Xu},
  \citenamefont {Chen}, \citenamefont {Nahas}, \citenamefont {Prokhorenko},\
  and\ \citenamefont {Bellaiche}}]{RN1059}%
  \BibitemOpen
  \bibfield  {author} {\bibinfo {author} {\bibfnamefont {Y.}~\bibnamefont
  {Zhang}}, \bibinfo {author} {\bibfnamefont {C.}~\bibnamefont {Xu}}, \bibinfo
  {author} {\bibfnamefont {P.}~\bibnamefont {Chen}}, \bibinfo {author}
  {\bibfnamefont {Y.}~\bibnamefont {Nahas}}, \bibinfo {author} {\bibfnamefont
  {S.}~\bibnamefont {Prokhorenko}}, \ and\ \bibinfo {author} {\bibfnamefont
  {L.}~\bibnamefont {Bellaiche}},\ }\href {\doibase
  10.1103/PhysRevB.102.241107} {\bibfield  {journal} {\bibinfo  {journal}
  {Phys. Rev. B}\ }\textbf {\bibinfo {volume} {102}},\ \bibinfo {pages}
  {241107(R)} (\bibinfo {year} {2020}{\natexlab{b}})}\BibitemShut {NoStop}%
\bibitem [{\citenamefont {Yuan}\ \emph
  {et~al.}(2020{\natexlab{a}})\citenamefont {Yuan}, \citenamefont {Yang},
  \citenamefont {Cai}, \citenamefont {Wu}, \citenamefont {Chen}, \citenamefont
  {Yan},\ and\ \citenamefont {Shen}}]{PhysRevB.101.094420}%
  \BibitemOpen
  \bibfield  {author} {\bibinfo {author} {\bibfnamefont {J.}~\bibnamefont
  {Yuan}}, \bibinfo {author} {\bibfnamefont {Y.}~\bibnamefont {Yang}}, \bibinfo
  {author} {\bibfnamefont {Y.}~\bibnamefont {Cai}}, \bibinfo {author}
  {\bibfnamefont {Y.}~\bibnamefont {Wu}}, \bibinfo {author} {\bibfnamefont
  {Y.}~\bibnamefont {Chen}}, \bibinfo {author} {\bibfnamefont {X.}~\bibnamefont
  {Yan}}, \ and\ \bibinfo {author} {\bibfnamefont {L.}~\bibnamefont {Shen}},\
  }\href {\doibase 10.1103/PhysRevB.101.094420} {\bibfield  {journal} {\bibinfo
   {journal} {Phys. Rev. B}\ }\textbf {\bibinfo {volume} {101}},\ \bibinfo
  {pages} {094420} (\bibinfo {year} {2020}{\natexlab{a}})}\BibitemShut
  {NoStop}%
\bibitem [{\citenamefont {Jiang}\ \emph {et~al.}(2021)\citenamefont {Jiang},
  \citenamefont {Liu}, \citenamefont {Li},\ and\ \citenamefont {Mi}}]{RN3528}%
  \BibitemOpen
  \bibfield  {author} {\bibinfo {author} {\bibfnamefont {J.}~\bibnamefont
  {Jiang}}, \bibinfo {author} {\bibfnamefont {X.}~\bibnamefont {Liu}}, \bibinfo
  {author} {\bibfnamefont {R.}~\bibnamefont {Li}}, \ and\ \bibinfo {author}
  {\bibfnamefont {W.}~\bibnamefont {Mi}},\ }\href {\doibase 10.1063/5.0057794}
  {\bibfield  {journal} {\bibinfo  {journal} {Appl. Phys. Lett.}\ }\textbf
  {\bibinfo {volume} {119}},\ \bibinfo {pages} {072401} (\bibinfo {year}
  {2021})}\BibitemShut {NoStop}%
\bibitem [{\citenamefont {Kresse}\ and\ \citenamefont
  {Furthmuller}(1996)}]{RN2004}%
  \BibitemOpen
  \bibfield  {author} {\bibinfo {author} {\bibfnamefont {G.}~\bibnamefont
  {Kresse}}\ and\ \bibinfo {author} {\bibfnamefont {J.}~\bibnamefont
  {Furthmuller}},\ }\href {\doibase 10.1103/PhysRevB.54.11169} {\bibfield
  {journal} {\bibinfo  {journal} {Phys. Rev. B}\ }\textbf {\bibinfo {volume}
  {54}},\ \bibinfo {pages} {11169} (\bibinfo {year} {1996})}\BibitemShut
  {NoStop}%
\bibitem [{\citenamefont {Kresse}\ and\ \citenamefont
  {Furthmüller}(1996)}]{RN1060}%
  \BibitemOpen
  \bibfield  {author} {\bibinfo {author} {\bibfnamefont {G.}~\bibnamefont
  {Kresse}}\ and\ \bibinfo {author} {\bibfnamefont {J.}~\bibnamefont
  {Furthmüller}},\ }\href {\doibase 10.1016/0927-0256(96)00008-0} {\bibfield
  {journal} {\bibinfo  {journal} {Comput. Mater. Sci.}\ }\textbf {\bibinfo
  {volume} {6}},\ \bibinfo {pages} {15} (\bibinfo {year} {1996})}\BibitemShut
  {NoStop}%
\bibitem [{\citenamefont {Kresse}\ and\ \citenamefont {Hafner}(1994)}]{RN1064}%
  \BibitemOpen
  \bibfield  {author} {\bibinfo {author} {\bibfnamefont {G.}~\bibnamefont
  {Kresse}}\ and\ \bibinfo {author} {\bibfnamefont {J.}~\bibnamefont
  {Hafner}},\ }\href {\doibase 10.1103/physrevb.49.14251} {\bibfield  {journal}
  {\bibinfo  {journal} {Phys. Rev. B Condens. Matter}\ }\textbf {\bibinfo
  {volume} {49}},\ \bibinfo {pages} {14251} (\bibinfo {year}
  {1994})}\BibitemShut {NoStop}%
\bibitem [{\citenamefont {Kresse}\ and\ \citenamefont {Hafner}(1993)}]{RN1061}%
  \BibitemOpen
  \bibfield  {author} {\bibinfo {author} {\bibfnamefont {G.}~\bibnamefont
  {Kresse}}\ and\ \bibinfo {author} {\bibfnamefont {J.}~\bibnamefont
  {Hafner}},\ }\href {\doibase 10.1103/physrevb.47.558} {\bibfield  {journal}
  {\bibinfo  {journal} {Phys. Rev. B Condens. Matter}\ }\textbf {\bibinfo
  {volume} {47}},\ \bibinfo {pages} {558} (\bibinfo {year} {1993})}\BibitemShut
  {NoStop}%
\bibitem [{\citenamefont {Perdew}\ \emph {et~al.}(1996)\citenamefont {Perdew},
  \citenamefont {Burke},\ and\ \citenamefont {Ernzerhof}}]{RN1999}%
  \BibitemOpen
  \bibfield  {author} {\bibinfo {author} {\bibfnamefont {J.~P.}\ \bibnamefont
  {Perdew}}, \bibinfo {author} {\bibfnamefont {K.}~\bibnamefont {Burke}}, \
  and\ \bibinfo {author} {\bibfnamefont {M.}~\bibnamefont {Ernzerhof}},\ }\href
  {\doibase 10.1103/PhysRevLett.77.3865} {\bibfield  {journal} {\bibinfo
  {journal} {Phys. Rev. Lett.}\ }\textbf {\bibinfo {volume} {77}},\ \bibinfo
  {pages} {3865} (\bibinfo {year} {1996})}\BibitemShut {NoStop}%
\bibitem [{\citenamefont {Dudarev}\ \emph {et~al.}(1998)\citenamefont
  {Dudarev}, \citenamefont {Botton}, \citenamefont {Savrasov}, \citenamefont
  {Humphreys},\ and\ \citenamefont {Sutton}}]{Dudarev1998a}%
  \BibitemOpen
  \bibfield  {author} {\bibinfo {author} {\bibfnamefont {S.~L.}\ \bibnamefont
  {Dudarev}}, \bibinfo {author} {\bibfnamefont {G.~A.}\ \bibnamefont {Botton}},
  \bibinfo {author} {\bibfnamefont {S.~Y.}\ \bibnamefont {Savrasov}}, \bibinfo
  {author} {\bibfnamefont {C.~J.}\ \bibnamefont {Humphreys}}, \ and\ \bibinfo
  {author} {\bibfnamefont {A.~P.}\ \bibnamefont {Sutton}},\ }\href {\doibase
  10.1103/PhysRevB.57.1505} {\bibfield  {journal} {\bibinfo  {journal} {Phys.
  Rev. B}\ }\textbf {\bibinfo {volume} {57}},\ \bibinfo {pages} {1505}
  (\bibinfo {year} {1998})}\BibitemShut {NoStop}%
\bibitem [{\citenamefont {Togo}\ \emph {et~al.}(2008)\citenamefont {Togo},
  \citenamefont {Oba},\ and\ \citenamefont {Tanaka}}]{Togo2008}%
  \BibitemOpen
  \bibfield  {author} {\bibinfo {author} {\bibfnamefont {A.}~\bibnamefont
  {Togo}}, \bibinfo {author} {\bibfnamefont {F.}~\bibnamefont {Oba}}, \ and\
  \bibinfo {author} {\bibfnamefont {I.}~\bibnamefont {Tanaka}},\ }\href
  {\doibase 10.1103/PhysRevB.78.134106} {\bibfield  {journal} {\bibinfo
  {journal} {Phys. Rev. B}\ }\textbf {\bibinfo {volume} {78}},\ \bibinfo
  {pages} {134106} (\bibinfo {year} {2008})}\BibitemShut {NoStop}%
\bibitem [{\citenamefont {Togo}\ and\ \citenamefont {Tanaka}(2015)}]{Togo2015}%
  \BibitemOpen
  \bibfield  {author} {\bibinfo {author} {\bibfnamefont {A.}~\bibnamefont
  {Togo}}\ and\ \bibinfo {author} {\bibfnamefont {I.}~\bibnamefont {Tanaka}},\
  }\href {\doibase https://doi.org/10.1016/j.scriptamat.2015.07.021} {\bibfield
   {journal} {\bibinfo  {journal} {Scr. Mater}\ }\textbf {\bibinfo {volume}
  {108}},\ \bibinfo {pages} {1} (\bibinfo {year} {2015})}\BibitemShut {NoStop}%
\bibitem [{\citenamefont {M\"uller}\ \emph {et~al.}(2019)\citenamefont
  {M\"uller}, \citenamefont {Hoffmann}, \citenamefont {Di\ss{}elkamp},
  \citenamefont {Sch\"urhoff}, \citenamefont {Mavros}, \citenamefont
  {Sallermann}, \citenamefont {Kiselev}, \citenamefont {J\'onsson},\ and\
  \citenamefont {Bl\"ugel}}]{RN1055}%
  \BibitemOpen
  \bibfield  {author} {\bibinfo {author} {\bibfnamefont {G.~P.}\ \bibnamefont
  {M\"uller}}, \bibinfo {author} {\bibfnamefont {M.}~\bibnamefont {Hoffmann}},
  \bibinfo {author} {\bibfnamefont {C.}~\bibnamefont {Di\ss{}elkamp}}, \bibinfo
  {author} {\bibfnamefont {D.}~\bibnamefont {Sch\"urhoff}}, \bibinfo {author}
  {\bibfnamefont {S.}~\bibnamefont {Mavros}}, \bibinfo {author} {\bibfnamefont
  {M.}~\bibnamefont {Sallermann}}, \bibinfo {author} {\bibfnamefont {N.~S.}\
  \bibnamefont {Kiselev}}, \bibinfo {author} {\bibfnamefont {H.}~\bibnamefont
  {J\'onsson}}, \ and\ \bibinfo {author} {\bibfnamefont {S.}~\bibnamefont
  {Bl\"ugel}},\ }\href {\doibase 10.1103/PhysRevB.99.224414} {\bibfield
  {journal} {\bibinfo  {journal} {Phys. Rev. B}\ }\textbf {\bibinfo {volume}
  {99}},\ \bibinfo {pages} {224414} (\bibinfo {year} {2019})}\BibitemShut
  {NoStop}%
\bibitem [{\citenamefont {Lifshitz}\ and\ \citenamefont {M.}(1935)}]{RN1984}%
  \BibitemOpen
  \bibfield  {author} {\bibinfo {author} {\bibfnamefont {L.~D.~L.}\
  \bibnamefont {Lifshitz}}\ and\ \bibinfo {author} {\bibfnamefont
  {E.}~\bibnamefont {M.}},\ }\href {\doibase
  10.1016/B978-0-08-036364-6.50008-9} {\bibfield  {journal} {\bibinfo
  {journal} {Phys. Z. Sowjetunion}\ }\textbf {\bibinfo {volume} {8}},\ \bibinfo
  {pages} {51} (\bibinfo {year} {1935})}\BibitemShut {NoStop}%
\bibitem [{\citenamefont {Gilbert}(2004)}]{RN1985}%
  \BibitemOpen
  \bibfield  {author} {\bibinfo {author} {\bibfnamefont {T.~L.}\ \bibnamefont
  {Gilbert}},\ }\href {\doibase 10.1109/tmag.2004.836740} {\bibfield  {journal}
  {\bibinfo  {journal} {IEEE Trans. Magn.}\ }\textbf {\bibinfo {volume} {40}},\
  \bibinfo {pages} {3443} (\bibinfo {year} {2004})}\BibitemShut {NoStop}%
\bibitem [{\citenamefont {Liu}\ \emph {et~al.}(2019)\citenamefont {Liu},
  \citenamefont {Ren}, \citenamefont {Xie}, \citenamefont {Cheng},
  \citenamefont {Liu}, \citenamefont {An}, \citenamefont {Qin},\ and\
  \citenamefont {Hu}}]{Liu2019}%
  \BibitemOpen
  \bibfield  {author} {\bibinfo {author} {\bibfnamefont {L.}~\bibnamefont
  {Liu}}, \bibinfo {author} {\bibfnamefont {X.}~\bibnamefont {Ren}}, \bibinfo
  {author} {\bibfnamefont {J.}~\bibnamefont {Xie}}, \bibinfo {author}
  {\bibfnamefont {B.}~\bibnamefont {Cheng}}, \bibinfo {author} {\bibfnamefont
  {W.}~\bibnamefont {Liu}}, \bibinfo {author} {\bibfnamefont {T.}~\bibnamefont
  {An}}, \bibinfo {author} {\bibfnamefont {H.}~\bibnamefont {Qin}}, \ and\
  \bibinfo {author} {\bibfnamefont {J.}~\bibnamefont {Hu}},\ }\href {\doibase
  https://doi.org/10.1016/j.apsusc.2019.02.203} {\bibfield  {journal} {\bibinfo
   {journal} {Appl. Surf. Sci.}\ }\textbf {\bibinfo {volume} {480}},\ \bibinfo
  {pages} {300} (\bibinfo {year} {2019})}\BibitemShut {NoStop}%
\bibitem [{\citenamefont {Liu}\ \emph {et~al.}(2018{\natexlab{b}})\citenamefont
  {Liu}, \citenamefont {Shi}, \citenamefont {Lu},\ and\ \citenamefont
  {Anantram}}]{RN2}%
  \BibitemOpen
  \bibfield  {author} {\bibinfo {author} {\bibfnamefont {J.}~\bibnamefont
  {Liu}}, \bibinfo {author} {\bibfnamefont {M.}~\bibnamefont {Shi}}, \bibinfo
  {author} {\bibfnamefont {J.}~\bibnamefont {Lu}}, \ and\ \bibinfo {author}
  {\bibfnamefont {M.~P.}\ \bibnamefont {Anantram}},\ }\href {\doibase
  10.1103/PhysRevB.97.054416} {\bibfield  {journal} {\bibinfo  {journal} {Phys.
  Rev. B}\ }\textbf {\bibinfo {volume} {97}},\ \bibinfo {pages} {054416}
  (\bibinfo {year} {2018}{\natexlab{b}})}\BibitemShut {NoStop}%
\bibitem [{\citenamefont {Shen}\ \emph {et~al.}(2021)\citenamefont {Shen},
  \citenamefont {Song}, \citenamefont {Xue}, \citenamefont {Wu}, \citenamefont
  {Wang},\ and\ \citenamefont {Zhong}}]{shen2021}%
  \BibitemOpen
  \bibfield  {author} {\bibinfo {author} {\bibfnamefont {Z.}~\bibnamefont
  {Shen}}, \bibinfo {author} {\bibfnamefont {C.}~\bibnamefont {Song}}, \bibinfo
  {author} {\bibfnamefont {Y.}~\bibnamefont {Xue}}, \bibinfo {author}
  {\bibfnamefont {Z.}~\bibnamefont {Wu}}, \bibinfo {author} {\bibfnamefont
  {J.}~\bibnamefont {Wang}}, \ and\ \bibinfo {author} {\bibfnamefont
  {Z.}~\bibnamefont {Zhong}},\ }\href@noop {} {\  (\bibinfo {year} {2021})},\
  \Eprint {http://arxiv.org/abs/2109.00723} {arXiv:2109.00723} \BibitemShut
  {NoStop}%
\bibitem [{\citenamefont {Guo}\ \emph {et~al.}(2020)\citenamefont {Guo},
  \citenamefont {Wang}, \citenamefont {Zhang}, \citenamefont {Yuan},
  \citenamefont {Ma},\ and\ \citenamefont {Wang}}]{Guo}%
  \BibitemOpen
  \bibfield  {author} {\bibinfo {author} {\bibfnamefont {Y.}~\bibnamefont
  {Guo}}, \bibinfo {author} {\bibfnamefont {B.}~\bibnamefont {Wang}}, \bibinfo
  {author} {\bibfnamefont {X.}~\bibnamefont {Zhang}}, \bibinfo {author}
  {\bibfnamefont {S.}~\bibnamefont {Yuan}}, \bibinfo {author} {\bibfnamefont
  {L.}~\bibnamefont {Ma}}, \ and\ \bibinfo {author} {\bibfnamefont
  {J.}~\bibnamefont {Wang}},\ }\href {\doibase
  https://doi.org/10.1002/inf2.12096} {\bibfield  {journal} {\bibinfo
  {journal} {InfoMat}\ }\textbf {\bibinfo {volume} {2}},\ \bibinfo {pages}
  {639} (\bibinfo {year} {2020})}\BibitemShut {NoStop}%
\bibitem [{\citenamefont {Li}\ and\ \citenamefont {Yang}(2019)}]{Li2019}%
  \BibitemOpen
  \bibfield  {author} {\bibinfo {author} {\bibfnamefont {X.}~\bibnamefont
  {Li}}\ and\ \bibinfo {author} {\bibfnamefont {J.}~\bibnamefont {Yang}},\
  }\href {\doibase https://doi.org/10.1021/jacs.8b11346} {\bibfield  {journal}
  {\bibinfo  {journal} {J. Am. Chem. Soc.}\ }\textbf {\bibinfo {volume}
  {141}},\ \bibinfo {pages} {109} (\bibinfo {year} {2019})}\BibitemShut
  {NoStop}%
\bibitem [{\citenamefont {Goodenough}(1955)}]{Goodenough1955}%
  \BibitemOpen
  \bibfield  {author} {\bibinfo {author} {\bibfnamefont {J.~B.}\ \bibnamefont
  {Goodenough}},\ }\href {\doibase 10.1103/PhysRev.100.564} {\bibfield
  {journal} {\bibinfo  {journal} {Phys. Rev.}\ }\textbf {\bibinfo {volume}
  {100}},\ \bibinfo {pages} {564} (\bibinfo {year} {1955})}\BibitemShut
  {NoStop}%
\bibitem [{\citenamefont {Kanamori}(1959)}]{Kanamori1959}%
  \BibitemOpen
  \bibfield  {author} {\bibinfo {author} {\bibfnamefont {J.}~\bibnamefont
  {Kanamori}},\ }\href {\doibase https://doi.org/10.1016/0022-3697(59)90061-7}
  {\bibfield  {journal} {\bibinfo  {journal} {J. Phys. Chem. Solids}\ }\textbf
  {\bibinfo {volume} {10}},\ \bibinfo {pages} {87} (\bibinfo {year}
  {1959})}\BibitemShut {NoStop}%
\bibitem [{\citenamefont {Anderson}(1959)}]{Anderson1959}%
  \BibitemOpen
  \bibfield  {author} {\bibinfo {author} {\bibfnamefont {P.~W.}\ \bibnamefont
  {Anderson}},\ }\href {\doibase 10.1103/PhysRev.115.2} {\bibfield  {journal}
  {\bibinfo  {journal} {Phys. Rev.}\ }\textbf {\bibinfo {volume} {115}},\
  \bibinfo {pages} {2} (\bibinfo {year} {1959})}\BibitemShut {NoStop}%
\bibitem [{\citenamefont {Lado}\ and\ \citenamefont
  {Fernandez-Rossier}(2017)}]{Lado2017}%
  \BibitemOpen
  \bibfield  {author} {\bibinfo {author} {\bibfnamefont {J.~L.}\ \bibnamefont
  {Lado}}\ and\ \bibinfo {author} {\bibfnamefont {J.}~\bibnamefont
  {Fernandez-Rossier}},\ }\href {\doibase 10.1088/2053-1583/aa75ed} {\bibfield
  {journal} {\bibinfo  {journal} {2D Mater.}\ }\textbf {\bibinfo {volume}
  {4}},\ \bibinfo {pages} {035002} (\bibinfo {year} {2017})}\BibitemShut
  {NoStop}%
\bibitem [{\citenamefont {Tartaglia}\ \emph {et~al.}(2020)\citenamefont
  {Tartaglia}, \citenamefont {Tang}, \citenamefont {Lado}, \citenamefont
  {Bahrami}, \citenamefont {Abramchuk}, \citenamefont {McCandless},
  \citenamefont {Doyle}, \citenamefont {Burch}, \citenamefont {Ran},
  \citenamefont {Chan},\ and\ \citenamefont {Tafti}}]{Lado2020SA}%
  \BibitemOpen
  \bibfield  {author} {\bibinfo {author} {\bibfnamefont {T.~A.}\ \bibnamefont
  {Tartaglia}}, \bibinfo {author} {\bibfnamefont {J.~N.}\ \bibnamefont {Tang}},
  \bibinfo {author} {\bibfnamefont {J.~L.}\ \bibnamefont {Lado}}, \bibinfo
  {author} {\bibfnamefont {F.}~\bibnamefont {Bahrami}}, \bibinfo {author}
  {\bibfnamefont {M.}~\bibnamefont {Abramchuk}}, \bibinfo {author}
  {\bibfnamefont {G.~T.}\ \bibnamefont {McCandless}}, \bibinfo {author}
  {\bibfnamefont {M.~C.}\ \bibnamefont {Doyle}}, \bibinfo {author}
  {\bibfnamefont {K.~S.}\ \bibnamefont {Burch}}, \bibinfo {author}
  {\bibfnamefont {Y.}~\bibnamefont {Ran}}, \bibinfo {author} {\bibfnamefont
  {J.~Y.}\ \bibnamefont {Chan}}, \ and\ \bibinfo {author} {\bibfnamefont
  {F.}~\bibnamefont {Tafti}},\ }\href {\doibase
  https://doi.org/10.1126/sciadv.abb9379} {\bibfield  {journal} {\bibinfo
  {journal} {Sci. Adv.}\ }\textbf {\bibinfo {volume} {6}},\ \bibinfo {pages}
  {eabb9379} (\bibinfo {year} {2020})}\BibitemShut {NoStop}%
\bibitem [{\citenamefont {Fert}\ and\ \citenamefont
  {Levy}(1980)}]{PhysRevLett.44.1538}%
  \BibitemOpen
  \bibfield  {author} {\bibinfo {author} {\bibfnamefont {A.}~\bibnamefont
  {Fert}}\ and\ \bibinfo {author} {\bibfnamefont {P.~M.}\ \bibnamefont
  {Levy}},\ }\href {\doibase 10.1103/PhysRevLett.44.1538} {\bibfield  {journal}
  {\bibinfo  {journal} {Phys. Rev. Lett.}\ }\textbf {\bibinfo {volume} {44}},\
  \bibinfo {pages} {1538} (\bibinfo {year} {1980})}\BibitemShut {NoStop}%
\bibitem [{\citenamefont {Berg}\ and\ \citenamefont
  {Lüscher}(1981)}]{Berg1981}%
  \BibitemOpen
  \bibfield  {author} {\bibinfo {author} {\bibfnamefont {B.}~\bibnamefont
  {Berg}}\ and\ \bibinfo {author} {\bibfnamefont {M.}~\bibnamefont
  {Lüscher}},\ }\href {\doibase https://doi.org/10.1016/0550-3213(81)90568-X}
  {\bibfield  {journal} {\bibinfo  {journal} {Nucl. Phys. B}\ }\textbf
  {\bibinfo {volume} {190}},\ \bibinfo {pages} {412} (\bibinfo {year}
  {1981})}\BibitemShut {NoStop}%
\bibitem [{\citenamefont {Yin}\ \emph {et~al.}(2016)\citenamefont {Yin},
  \citenamefont {Li}, \citenamefont {Kong}, \citenamefont {Lake}, \citenamefont
  {Chien},\ and\ \citenamefont {Zang}}]{PhysRevB.93.174403}%
  \BibitemOpen
  \bibfield  {author} {\bibinfo {author} {\bibfnamefont {G.}~\bibnamefont
  {Yin}}, \bibinfo {author} {\bibfnamefont {Y.}~\bibnamefont {Li}}, \bibinfo
  {author} {\bibfnamefont {L.}~\bibnamefont {Kong}}, \bibinfo {author}
  {\bibfnamefont {R.~K.}\ \bibnamefont {Lake}}, \bibinfo {author}
  {\bibfnamefont {C.~L.}\ \bibnamefont {Chien}}, \ and\ \bibinfo {author}
  {\bibfnamefont {J.}~\bibnamefont {Zang}},\ }\href {\doibase
  10.1103/PhysRevB.93.174403} {\bibfield  {journal} {\bibinfo  {journal} {Phys.
  Rev. B}\ }\textbf {\bibinfo {volume} {93}},\ \bibinfo {pages} {174403}
  (\bibinfo {year} {2016})}\BibitemShut {NoStop}%
\bibitem [{\citenamefont {Rozsa}\ \emph {et~al.}(2016)\citenamefont {Rozsa},
  \citenamefont {Simon}, \citenamefont {Palotas}, \citenamefont {Udvardi},\
  and\ \citenamefont {Szunyogh}}]{RN3124}%
  \BibitemOpen
  \bibfield  {author} {\bibinfo {author} {\bibfnamefont {L.}~\bibnamefont
  {Rozsa}}, \bibinfo {author} {\bibfnamefont {E.}~\bibnamefont {Simon}},
  \bibinfo {author} {\bibfnamefont {K.}~\bibnamefont {Palotas}}, \bibinfo
  {author} {\bibfnamefont {L.}~\bibnamefont {Udvardi}}, \ and\ \bibinfo
  {author} {\bibfnamefont {L.}~\bibnamefont {Szunyogh}},\ }\href {\doibase
  10.1103/PhysRevB.93.024417} {\bibfield  {journal} {\bibinfo  {journal} {Phys.
  Rev. B}\ }\textbf {\bibinfo {volume} {93}},\ \bibinfo {pages} {024417}
  (\bibinfo {year} {2016})}\BibitemShut {NoStop}%
\bibitem [{\citenamefont {Du}\ \emph {et~al.}(2022)\citenamefont {Du},
  \citenamefont {Dou}, \citenamefont {He}, \citenamefont {Dai}, \citenamefont
  {Huang},\ and\ \citenamefont {Ma}}]{RN3705}%
  \BibitemOpen
  \bibfield  {author} {\bibinfo {author} {\bibfnamefont {W.}~\bibnamefont
  {Du}}, \bibinfo {author} {\bibfnamefont {K.}~\bibnamefont {Dou}}, \bibinfo
  {author} {\bibfnamefont {Z.}~\bibnamefont {He}}, \bibinfo {author}
  {\bibfnamefont {Y.}~\bibnamefont {Dai}}, \bibinfo {author} {\bibfnamefont
  {B.}~\bibnamefont {Huang}}, \ and\ \bibinfo {author} {\bibfnamefont
  {Y.}~\bibnamefont {Ma}},\ }\href {\doibase 10.1021/acs.nanolett.2c00836}
  {\bibfield  {journal} {\bibinfo  {journal} {Nano Lett.}\ }\textbf {\bibinfo
  {volume} {22}},\ \bibinfo {pages} {3440} (\bibinfo {year}
  {2022})}\BibitemShut {NoStop}%
\bibitem [{\citenamefont {Yuan}\ \emph
  {et~al.}(2020{\natexlab{b}})\citenamefont {Yuan}, \citenamefont {Yang},
  \citenamefont {Cai}, \citenamefont {Wu}, \citenamefont {Chen}, \citenamefont
  {Yan},\ and\ \citenamefont {Shen}}]{RN1042}%
  \BibitemOpen
  \bibfield  {author} {\bibinfo {author} {\bibfnamefont {J.}~\bibnamefont
  {Yuan}}, \bibinfo {author} {\bibfnamefont {Y.}~\bibnamefont {Yang}}, \bibinfo
  {author} {\bibfnamefont {Y.}~\bibnamefont {Cai}}, \bibinfo {author}
  {\bibfnamefont {Y.}~\bibnamefont {Wu}}, \bibinfo {author} {\bibfnamefont
  {Y.}~\bibnamefont {Chen}}, \bibinfo {author} {\bibfnamefont {X.}~\bibnamefont
  {Yan}}, \ and\ \bibinfo {author} {\bibfnamefont {L.}~\bibnamefont {Shen}},\
  }\href {\doibase 10.1103/PhysRevB.101.094420} {\bibfield  {journal} {\bibinfo
   {journal} {Phys. Rev. B}\ }\textbf {\bibinfo {volume} {101}},\ \bibinfo
  {pages} {094420} (\bibinfo {year} {2020}{\natexlab{b}})}\BibitemShut
  {NoStop}%
\bibitem [{\citenamefont {Zeng}\ \emph {et~al.}(2020)\citenamefont {Zeng},
  \citenamefont {Zhang}, \citenamefont {Jin}, \citenamefont {Wang},
  \citenamefont {Song}, \citenamefont {Ma}, \citenamefont {Liu},\ and\
  \citenamefont {Wang}}]{RN3663}%
  \BibitemOpen
  \bibfield  {author} {\bibinfo {author} {\bibfnamefont {Z.}~\bibnamefont
  {Zeng}}, \bibinfo {author} {\bibfnamefont {C.}~\bibnamefont {Zhang}},
  \bibinfo {author} {\bibfnamefont {C.}~\bibnamefont {Jin}}, \bibinfo {author}
  {\bibfnamefont {J.}~\bibnamefont {Wang}}, \bibinfo {author} {\bibfnamefont
  {C.}~\bibnamefont {Song}}, \bibinfo {author} {\bibfnamefont {Y.}~\bibnamefont
  {Ma}}, \bibinfo {author} {\bibfnamefont {Q.}~\bibnamefont {Liu}}, \ and\
  \bibinfo {author} {\bibfnamefont {J.}~\bibnamefont {Wang}},\ }\href {\doibase
  10.1063/5.0022527} {\bibfield  {journal} {\bibinfo  {journal} {Appl. Phys.
  Lett.}\ }\textbf {\bibinfo {volume} {117}},\ \bibinfo {pages} {172404}
  (\bibinfo {year} {2020})}\BibitemShut {NoStop}%
\bibitem [{\citenamefont {Everschor-Sitte}\ \emph {et~al.}(2018)\citenamefont
  {Everschor-Sitte}, \citenamefont {Masell}, \citenamefont {Reeve},\ and\
  \citenamefont {Klaeui}}]{RN3063}%
  \BibitemOpen
  \bibfield  {author} {\bibinfo {author} {\bibfnamefont {K.}~\bibnamefont
  {Everschor-Sitte}}, \bibinfo {author} {\bibfnamefont {J.}~\bibnamefont
  {Masell}}, \bibinfo {author} {\bibfnamefont {R.~M.}\ \bibnamefont {Reeve}}, \
  and\ \bibinfo {author} {\bibfnamefont {M.}~\bibnamefont {Klaeui}},\ }\href
  {\doibase 10.1063/1.5048972} {\bibfield  {journal} {\bibinfo  {journal} {J.
  Appl. Phys.}\ }\textbf {\bibinfo {volume} {124}},\ \bibinfo {pages} {240901}
  (\bibinfo {year} {2018})}\BibitemShut {NoStop}%
\end{thebibliography}%
\end{document}